\documentclass[twocolumn]{aastex631}

\usepackage{physics}
\usepackage[caption=false]{subfig}
\shorttitle{Kinetic simulations of CRPAI}
\shortauthors{X. Sun, X.-N. Bai and X. Zhao}

\begin{document}

\title{Kinetic simulations of the cosmic ray pressure anisotropy instability: cosmic ray scattering rate in the saturated state}

\correspondingauthor{Xiaochen Sun, Xue-Ning Bai}

\author[0009-0009-7676-6188]{Xiaochen Sun}
\affiliation{Institute for Advanced Study, Tsinghua University, Beijing 100084, China}
\affiliation{Department of Astrophysical Sciences, Princeton University, Princeton, NJ 08544, USA}
\email{sun.xiaochen@princeton.edu}

\author[0000-0001-6906-9549]{Xue-Ning Bai}
\affiliation{Institute for Advanced Study, Tsinghua University, Beijing 100084, China}
\affiliation{Department of Astronomy, Tsinghua University, Beijing 100084, China}
\email{xbai@tsinghua.edu.cn}

\author[0009-0003-1259-319X]{Xihui Zhao}
\affiliation{Institute for Advanced Study, Tsinghua University, Beijing 100084, China}



\begin{abstract}
Cosmic ray (CR) feedback plays a vital role in shaping the formation and evolution of galaxies through their interaction with magnetohydrodynamic waves. In the CR self-confinement scenario, the waves are generated by the CR gyro-resonant instabilities via CR streaming or CR pressure anisotropy and saturate by balancing wave damping. The resulting effective particle scattering rate by the waves, $\nu_{\text{eff}}$, critically sets the coupling between the CRs and background gas, but the efficiency of CR feedback is yet poorly constrained. We employ 1D kinetic simulations under the Magnetohydrodynamic-Particle-In-Cell (MHD-PIC) framework with the adaptive $\delta f$ method to quantify $\nu_{\text{eff}}$ for the saturated state of the CR pressure anisotropy instability (CRPAI) with ion-neutral friction. We drive CR pressure anisotropy by expanding/compressing box, mimicking the background evolution of magnetic field strength, and the CR pressure anisotropy eventually reaches a quasi-steady state by balancing quasi-linear diffusion. At the saturated state, we measure $\nu_{\text{eff}}$ and the CR pressure anisotropy level, establishing a calibrated scaling relation with environmental parameters. The scaling relation is consistent with quasi-linear theory and can be incorporated to CR fluid models, in either the single-fluid or $p$-by-$p$ treatments. Our results serve as a basis for accurately calibrating the subgrid physics in macroscopic studies of CR feedback and transport.
\end{abstract}


\section{Introduction}
Cosmic rays (CRs) are (trans-)relativistic charged particles pervading in space. With relatively low number density ($\sim 10^{-9} \text{cm}^{-3}$), they possess an energy density ($\geq 1 \text{eV cm}^{-3}$) comparable to the internal, kinetic and magnetic energy density of gas in the Galaxy \citep[see reviews, e.g. ][]{2013A&ARv..21...70B,2013PhPl...20e5501Z, 2015ARA&A..53..199G}. Consequently, CRs are expected to play a dynamically important role, known as CR feedback \citep[see reviews, e.g.,][]{1997AdSpR..19..697P, 2001RvMP...73.1031F,2017PhPl...24e5402Z, 2017ARA&A..55...59N}. At a more fundamental level, the interaction between the CRs and the (ionized) gas is primarily through fluctuations (waves) in the background electromagnetic field \citep[e.g.,][]{1966ApJ...146..480J, 1975RvGSP..13..547V}, as opposed to direct collisions \citep[e.g.,][]{1974PhRvC...9.1718R, 1988PhRvC..37.1490F, 1989ApJ...336..243S}. Based on the source of such waves/fluctuations, CR feedback/transport are classified into two categories: fluctuations generated by CRs themselves, known as ``self-confinement" \citep[e.g.,][]{1969ApJ...156..303W, 1971ApJ...170..265S, 1979ApJ...228..576H}, and fluctuations cascaded from extrinsic turbulence \citep[e.g.,][]{1941DoSSR..30..301K, 1964SvA.....7..566I, 1965PhFl....8.1385K, 1995ApJ...438..763G}. As CR scattering and transport are primarily sensitive to electromagnetic fluctuations at their gyration scale, it is expected that self-confinement governs low-energy CRs (from sub-$\text{GeV}$ to $\sim 10^2 \text{GeV}$, \citealp{2012PhRvL.109f1101B, 2018PhRvL.121b1102E}), while extrinsic turbulence likely dominates the transport of higher-energy CRs.

As the CR energy density typically peaks around $\sim\text{GeV}$ \citep[e.g.,][]{2011Sci...332...69A, 2013ApJ...765...91A, 2014arXiv1402.0467C}, self-confinement is expected to be the dominant process driving the force of CR feedback. At microphysical level, this is mediated by the CR gyro-resonant instabilities \citep[e.g.,][]{1967ApJ...147..689L, 1968ApJ...152..987W, 1969ApJ...156..445K, 1973Ap&SS..21...13G, 1990acr..book.....B}. Here, magnetohydrodynamic (MHD) waves in the background gas can be destabilized when the CRs become weakly anisotropic with respect to the bulk gas (by more than about $U_A/c$, where $U_A$ is the Alfv\'en speed) through gyro-resonant interactions between the CRs and the waves. The classic flavor is the CR streaming instability \citep[CRSI, e.g. ][]{1974ARA&A..12...71W, 1975MNRAS.173..255S, 1975MNRAS.173..245S, 1975MNRAS.172..557S}, which occurs when the speed of CR bulk motion relative to background gas exceeds $U_A$. This situation naturally arises when the CRs escape from the source. With the gyro-resonant interactions, the growth of MHD waves is fed by the free energy from CR streaming, which in turn reduces the bulk CR motion towards the Alfv\'en wave speed \citep{2018JPlPh..84a9007A}. Another flavor of the CR gyro-resonant instability, which is poorly studied, is the CR pressure anisotropy instability \citep[CRPAI, e.g. ][]{2006MNRAS.373.1195L, 2011ApJ...731...35Y, 2020ApJ...890...67Z}, which arises when the level of CR pressure anisotropy exceeds $\sim U_A/c$. This situation naturally occurs when the bulk gas undergoes compression/expansion and/or shear, where CR anisotropy is driven by a change in background field strength $\dot{B}/B$ under the conservation of magnetic moment. The interaction between CRs and electromagnetic fluctuations tends to isotropize the CR momentum distribution to within the $\sim U_A/c$ level in the gas co-moving frame.

The coupling between the CRs and the background gas is reflected in the effective scattering rate $\nu_{\rm eff}$, which is determined by the amplitude of waves. In the self-confinement regime, the wave amplitudes are set by the balance between driving (from the CR gyro-resonant instabilities) and various wave damping mechanisms, which eventually determines the efficiency of CR feedback. These include the ion-neutral collisions \citep{1969ApJ...156..445K, 2016A&A...592A..28S}, the non-linear Landau damping \citep{1973Ap&SS..24...31L}, the linear Landau damping \citep{1979ApJ...233..302F, 2018MNRAS.473.3095W}, the extrinsic turbulence \citep{2004ApJ...604..671F, 2016ApJ...833..131L, 2022ApJ...927...94X}, and the influence of charged dust grains \citep{2021MNRAS.502.2630S}.  Which of the mechanisms would dominate depends on the local environment, with ion-neutral damping dominating in the neutral medium, while non-linear Landau damping is more prevalent in dilute gas \citep{2013PhPl...20e5501Z, 2021ApJ...922...11A, 2022ApJ...929..170A, 2024arXiv240104169A}.

At macroscopic level, galaxy simulations often model CRs as a fluid and employ approximate prescriptions for the effective scattering rate. Modeled as fluids instead of individual kinetic particles, the CR population is characterized by its energy density, momentum flux, and pressure \citep{1982A&A...116..191M, 2008MNRAS.384..251G, 2017MNRAS.465.4500P, 2018ApJ...854....5J, 2019MNRAS.485.2977T, 2019MNRAS.488.3716C}. Early simulations at galactic scales typically adopt highly simplified treatment (e.g., constant diffusion or streaming) of CR scattering, generally finding significant dynamical consequences especially in heating and in driving large-scale galactic outflows \citep[e.g.,][]{2008MNRAS.384..251G, 2012A&A...540A..77D, 2012MNRAS.423.2374U, 2013ApJ...777L..38H, 2016ApJ...816L..19G, 2017MNRAS.467..906W, 2018MNRAS.475..570J, 2018ApJ...868..108B, 2019MNRAS.488.3716C, 2019A&A...631A.121D, 2020A&A...638A.123D}, but the results are highly sensitive to the CR prescriptions. More recent works start to incorporate more realistic prescriptions of the CR scattering rates based on the CRSI and various damping mechanisms \citep{2021ApJ...922...11A,2024arXiv240104169A}, emphasizing the importance of resolving the multi-phase medium, while current state-of-the-art models are still unable to reproduce all observational constraints from the Galaxy \citep{2021MNRAS.501.4184H}.

The main motivation of this work is two-fold. First, we note that the scattering rates in the self-confinement regime, as calculated from the balance between wave growth and damping, are usually obtained under the quasi-linear theory (QLT) as in the aforementioned simulations. However, it involves several approximations that likely yield results that deviate from reality, and kinetic simulations are needed to clarify and calibrate these results \cite[e.g.][]{2022ApJ...928..112B}. Second, nearly all calculations so far focus on the CRSI \citep[e.g. ][]{2019ApJ...882....3H, 2019ApJ...876...60B, 2019ICRC...36..279H,2024arXiv240604400L}, which indeed likely dominates CR scattering and heating in the general situations \citep{2020ApJ...890...67Z}, but the role CRPAI have not been systematically investigated.
\citet{2018MNRAS.476.2779L} first studied the CRPAI numerically, starting from an initial high anisotropic CR distribution. They confirmed the linear growth of the CRPAI, but the wave amplitudes saturate following the isotropization of the CRs, set by the artificial initial condition. As one step forward, our recent work \citep{2023MNRAS.523.3328S} considered driving the CRPAI from an initially isotropic CR distribution. 
In this work, we simulate the saturation of CRPAI towards more realistic regimes, by continuously driving the CR anisotropy while incorporating wave damping, ensuring a steady-state balance at saturation. By combining kinetic simulations with analytical derivations, we aim to calibrate the CR scattering rates from the CRPAI from first principles.

Our simulations employ the magnetohydrodynamic-particle-in-cell (MHD-PIC) method \citep{1986JCoPh..66..469Z, 2000MNRAS.314...65L, 2015ApJ...809...55B}, which treats CRs kinetically as particles while treating the background plasma as a fluid described by MHD. Compared to the conventional PIC methods, it avoids resolving the microscopic scales of the background plasma, and hence becomes highly advantageous in simulating the CR gyro-resonant instabilities. Using the ATHENA MHD code \citep{2008ApJS..178..137S, 2015ApJ...809...55B}, \citet{2019ApJ...876...60B} developed the framework for simulating the CRSI, which captures the linear growth over a wide range of wavelengths and quasi-linear evolution over a broad range of CR energies around the peak of the energy distribution (which can be taken to be $\sim \text{GeV}$). This is facilitated by the implementation of the $\delta f$ method \citep[e.g. ][]{1993PhFlB...5...77P, 2014JCoPh.259..154K} to suppress Poisson noise, allowing one to use a reasonable number of particles to properly sample a weakly anisotropic particle distribution. Subsequent works of \citet{2021ApJ...914....3P, 2021ApJ...920..141B} further incorporated a simple prescription of ion-neutral damping. In \citet{2022ApJ...928..112B}, we introduced the streaming box framework, where CRs move along an imposed CR gradient, allowing the CRSI to be continuously driven, and it is balanced by wave damping. This framework thus enables us to measure the CR scattering rates at the saturated states from first principles. 

In this work, we migrate the MHD-PIC simulation to the \texttt{ATHENA++} MHD code \citep{2020ApJS..249....4S, 2023MNRAS.523.3328S}, which is more flexible and computationally efficient. We have already tested that the code well captures the linear growth of the CRPAI. In preparation for this work, we have also developed the expanding/compressing box \citep{1993PhRvL..70.2190G, 2015ApJ...800...88S, 2021ApJ...922L..35B} framework for MHD-PIC \citep{2023MNRAS.523.3328S}, which continuously drives the CR pressure anisotropy to trigger the CRPAI. This approach may mimic a local patch in compressible turbulence, or a mixing layer between two gas phases, or perhaps also including background shear, where a background changing field drives CR pressure anisotropy. In this work, we employ the expanding/compressing box to drive the CRPAI, balanced by ion-neutral damping to achieve a steady state. This allows us to quantitatively calibrate the effective CR scattering rate in the alternative scenario of CRPAI, establishing their dependence on local medium properties. This study thus fills a major gap in understanding the microphysics of CR feedback.

This paper is structured as follows: Section~\ref{sec:method} outlines the formulation and numerical setup. In particular, we introduce the adaptive $\delta f$ method in Section~\ref{sec::adapt_delta_f}. We calculate the expected CR scattering rates and their scaling relation under our simulation setup based on QLT in Section~\ref{sec::theory}. The simulation results are presented in Section~\ref{sec::result}, which are compared to quasi-linear theory for quantitative validation and calibration.
We discuss the implications of the results towards realistic systems in Section~\ref{sec::discuss}. In Section~\ref{sec::summary}, we conclude and discuss prospects for future research.

\section{Formulation and numerical methods}
\label{sec:method} 

In this section, we briefly review the MHD-PIC formulation in the expanding box framework \citep{2023MNRAS.523.3328S}, and the treatment of ion-neutral damping \citep{2021ApJ...914....3P, 2021ApJ...920..141B, 2022ApJ...928..112B}. We next propose a novel {\it adaptive} $\delta f$ method to reduce noise in our simulations, before describing our simulation setup.

\subsection{The governing equations}
\label{sec::equation}

The MHD-PIC method treats CRs as super-particles, following the standard approach of PIC, and solves MHD equations for the thermal gas (here specifically just for the thermal ion plasma, see later in this subsection). Each simulation particle carries both position information $\boldsymbol{x}$ and normalized momentum information $\left(\boldsymbol{p}/m\right)$. The CR kinetic equations are given as:
\begin{align}
    \frac{\dd \boldsymbol{x}}{\dd t} &= \boldsymbol{v}, \\
    \frac{\dd \left(\boldsymbol{p} / m\right)}{\dd t} &= \left(\frac{q}{mc}\right) \left(c \boldsymbol{E} + \boldsymbol{v} \times \boldsymbol{B} \right), \\
	\boldsymbol{v} &\equiv \frac{\left(\boldsymbol{p} / m\right)}{\gamma}, \quad \gamma \equiv \sqrt{1 + \left(\boldsymbol{p} / m\right)^2 / \mathbb{C}^2}.
\end{align}
Here, $q /\left(mc\right)$, $\boldsymbol{v}$, $\gamma$ and $\mathbb{C}$ represent the CR charge-to-mass ratio, particle velocity, Lorentz factor and the numerical speed of light, respectively. Here the numerical speed of light $\mathbb{C}$ sets the upper limit of the particle velocity in the simulation, and can be rescaled to match the physical speed of light $c$ in the realistic system (see Section~\ref{sec::theory_saturation}).
The CR population meanwhile exerts a Lorentz force on the MHD gas as CR backreaction. The MHD equations with the CR backreaction read,
\begin{align}
    \partial_t \rho_{\text{i}} + \nabla \cdot \left(\rho \boldsymbol{u}_{\text{i}}\right)= 0, \notag  \\
    \partial_t \left(\rho_{\text{i}} \boldsymbol{u}_{\text{i}}\right) + \nabla \cdot \left(\rho_{\text{i}} \boldsymbol{u}^T_{\text{i}} \boldsymbol{u}_{\text{i}} - \boldsymbol{B}^T \boldsymbol{B} + \mathbb{P}\right) \notag \\
    = - \nu_{\text{IN}}\rho_{\text{i}} \boldsymbol{u}_{\text{i}}
    -\left(\frac{Q_{\text{CR}}}{c} c\boldsymbol{E} + \frac{\boldsymbol{j}_{\text{CR}}}{c} \times \boldsymbol{B} \right), \notag \\
    \partial_t \boldsymbol{B} =-\nabla \times (c\boldsymbol{E}),\notag  \\
    c\boldsymbol{E} = -\boldsymbol{u}_{\text{i}} \times \boldsymbol{B},
\label{eq:mhd0}
\end{align}
Here, the variables include $\rho_{\text{i}}$ for ion gas density, $\mathbb{P}\equiv(\rho_{\text{i}} c_s^2 + B^2/2)\mathbb{I}$ for total pressure with $\mathbb{I}$ being the identity tensor, $c_s$ for the isothermal sound speed, and $\boldsymbol{u}_{\text{i}}$ for the ion gas velocity. The unit for magnetic field is normalized such that magnetic permeability equals one, thus absorbing factors of $1/\sqrt{4\pi}$. We bypass the CR backreaction on the MHD gas energy in this work by applying an isothermal equation of state (see the reason in Section~\ref{sec::theory_crhd}). The cosmic ray charge density $Q_{\text{CR}}$ and the current density $\boldsymbol{j}_{\text{CR}}$ are both functions of the phase space distribution function $f\left(t, \boldsymbol{x}, \boldsymbol{p}/m\right)$ for the CR population:
\begin{align}
    \frac{Q_{\text{CR}}}{c} \equiv & \left(\frac{q}{mc}\right) m \int f \dd^3 \boldsymbol{p} = \left(\frac{q}{mc}\right) \rho_{\text{CR}}, \notag  \\
    \frac{\boldsymbol{j}_{\text{CR}}}{c} \equiv &\left(\frac{q}{mc}\right) m \int \boldsymbol{v} f \dd^3 \boldsymbol{p}, \label{equ::feedback_define}
\end{align}
Here, $m$ represents the mass of an individual CR particle, and $\rho_{\text{CR}}$ denotes the mass density of the CR ensemble.

Given that the Lorentz force is approximately proportional to $\rho_{\text{CR}}$, the intensity of the CR backreaction is evaluated through the mass density ratio between CR and thermal ionized gas, $\rho_{\text{CR}}/\rho_{\text{i}}$. With the CR cyclotron frequency being $\Omega \equiv \left(q /\left(mc\right)\right) \left|\boldsymbol{B}\right|$, the characteristic rate for CR feedback is $(\rho_{\text{CR}}/\rho_{\text{i}})\Omega$.

We consider the ion-neutral friction as the wave damping mechanism in this work. Ion-neutral damping prevails as the primary damping mechanism in the cold dense medium \citep{2013PhPl...20e5501Z, 2021ApJ...922...11A, 2022ApJ...929..170A, 2022MNRAS.517.5413H}, where the ionization fraction $\rho_{\text{i}}/\rho_{\text{neu}} \ll 1$. Under typical ISM conditions, the ion-neutral damping can be treated such that we essentially only simulate the ion gas subjecting to a frictional force $- \nu_{\text{IN}}\rho_{\text{i}} \boldsymbol{u}_{\text{i}}$ on top of the background static neutrals \citep{2021ApJ...914....3P}, as included in Equation~\ref{eq:mhd0}. Here $\nu_{\text{IN}}$ represents the collision frequency of neutrals with ions. This approach remains valid only in the short-wavelength regime, where the Alfv\'en wave frequency $\omega \left(k\right) = k U_A$ for the relevant wavenumber $k$ exceeds $\nu_{\text{IN}}$ \citep{2007A&A...475..435R, 2021ApJ...914....3P}. Here, $U_A \equiv B / \sqrt{\rho_{\text{i}}}$ stands for the Alfv\'en velocity. For the $\sim \text{GeV}$ CRs we focus on in this work, the resonant wavenumber $k_\text{res}\sim \Omega m/p$, varies around $\sim \text{AU}^{-1}$, much shorter than $\nu_{\text{IN}} / U_A$ in typical ISM conditions \citep{2021ApJ...914....3P}, justifying our approach.
We also note that the induction equation retains its ideal MHD form.

In this work, we drive CR pressure anisotropy by mimicking a local uniform patch of gas among the macroscopic system (e.g., ISM/CGM, etc.) undergoing expansion/compression. This is achieved under the expanding box framework \citep[see derivations in ][]{1993PhRvL..70.2190G, 2015ApJ...800...88S,2023MNRAS.523.3328S}, and CR pressure anisotropy is developed owing to the conservation of the magnetic moment of individual CR particles. This framework has been implemented in the most general manner in the \texttt{ATHENA++} MHD-PIC module as described in \citet{2023MNRAS.523.3328S}, where the background expansion rate can be flexible in any of the three directions. In the simulations presented here, the background magnetic field lies along the $x$-direction. We let the background expand or compress at the rate $a(t)$ in the $y$- and $z$-directions and at a rate of $a^2(t)$ in the $x$-direction. This configuration enables MHD perturbations to propagate freely along the background magnetic field line \citep[see Appendix of ][]{2023MNRAS.523.3328S}. 

We reformulate the MHD equations and the CR equations of motion in comoving coordinates $\boldsymbol{x}'$ of the expanding box while expressing the MHD gas quantities and CR momentum in the lab frame. The equations we eventually solve in this study are as follows:
\begin{align}
	\partial_t \rho_{\text{i}} + \nabla' \cdot \left(\rho \boldsymbol{u}_{\text{i}}\right) &= -4 \rho_{\text{i}} \frac{\dot{a}}{a}, \label{equ::mass_conserve}  \\ 
	\partial_t \left(\rho_{\text{i}} \boldsymbol{u}_{\text{i}}\right) +& \nabla' \cdot \left(\rho_{\text{i}} \boldsymbol{u}^T_{\text{i}} \boldsymbol{u}_{\text{i}} - \boldsymbol{B}^T \boldsymbol{B} + \mathbb{P}\right) \notag \\ 
    &= -\nu_{\text{IN}}\rho_{\text{i}} \boldsymbol{u}_{\text{i}} -\left(\frac{Q_{\text{CR}}}{c} c\boldsymbol{E} + \frac{\boldsymbol{j}_{\text{CR}}}{c} \times \boldsymbol{B} \right) \notag \\ 
    &\quad - 4 \frac{\partial_t a}{a}\rho_{\text{i}} \boldsymbol{u}_{\text{i}} - \rho_{\text{i}} \boldsymbol{u}_{\text{i}} \cdot \mathbb{D}, \\
	\nabla' \cdot \boldsymbol{B} &= 0, \label{equ::gauss} \\
	\partial_{t} \boldsymbol{B}  - \nabla' \cross (c\boldsymbol{E}) &= - 4 \frac{\partial_t a}{a}\boldsymbol{B} + \mathbb{D} \cdot \boldsymbol{B}, \label{equ::induction} \\
	c\boldsymbol{E} &= -\boldsymbol{u}_{\text{i}} \times \boldsymbol{B}, \label{equ::frozen_in} \\
	\mathbb{A} \cdot \frac{\dd \boldsymbol{x}'}{\dd t} &= \boldsymbol{v},  \\
	\frac{\dd \left(\boldsymbol{p} / m\right)}{\dd t} + \mathbb{D} \cdot \frac{\boldsymbol{p}}{m} &= \frac{q}{mc} \left(c \boldsymbol{E} + \boldsymbol{v} \times \boldsymbol{B} \right), \label{equ::pic} 
\end{align}
where, 
\begin{align*}
	\nabla' &= \left(\frac{\partial}{a^2 \partial x'} ,\frac{\partial}{a \partial y'}, \frac{\partial}{a\partial z'} \right), \\
	\mathbb{D} &= diag\left(2 \frac{\partial_t a}{a}, \frac{\partial_t a}{a},\frac{\partial_t a}{a}\right),\\
	\mathbb{A} &= diag \left(a^2, a, a\right).
\end{align*} 
We note that even we use an isothermal equation of state, the background expansion or compression brings about a thermodynamic impact on the background MHD gas, following an adiabatic polytropic process \citep{2020ApJ...891L...2S}, $c_s \propto a^{-4/3} \left(t\right)$. The expansion (compression) dynamically reduces (amplifies) the background magnetic field at a rate of $\dot{B}/B = -2 \dot{a} / a$, and the time evolution of the background field strength is
\begin{equation*}
    B_g \left(t\right) = B_g \left( 0\right) \exp \left(-2 \frac{\dot{a}}{a} t\right).
\end{equation*}
Consequently, the CR cyclotron frequency $\Omega$ slowly varies over time, \textbf{$\Omega = \Omega(t=0) \exp \left(-2 \frac{\dot{a}}{a} t\right)$}. Both the gas density and CR density in the comoving frame change at a rate of $-4 \dot{a} / a$ with time. On the other hand, the mass density ratio $\rho_{\text{CR}}/\rho_{\text{i}}$, and the Alfv\'en speed $U_A$ remains constant throughout.

\subsection{Adaptive $\delta f$ method}
\label{sec::adapt_delta_f}
Inherent to a particle-based representation of the CR distribution function $f$, the MHD-PIC method is subject to Poisson noise, which represents a major issue in the simulation of the CR gyro-resonant instabilities \citep{2019ApJ...876...60B}. Simply increasing the particle number $N$ becomes unfeasible as the noise level only reduces as $N^{-1/2}$ which is very inefficient. This difficulty can be overcome by the $\delta f$ method \citep[e.g.][]{1993PhFlB...5...77P, 1994PhPl....1..863H, 1995JCoPh.119..283D, 2014JCoPh.259..154K, 2019ApJ...876...60B}.
In the $\delta f$ method, the phase space distribution $f\left(t, \boldsymbol{x}, \boldsymbol{p}/m\right)$ is divided into two components: $f_0\left(t, \boldsymbol{x}, \boldsymbol{p}/m\right)$ as a known analytical form, and $\delta f \left(t, \boldsymbol{x}, \boldsymbol{p}/m\right)$, representing the deviation $\delta f \equiv f-f_0$. Consequently, the CR backreaction also splits into two contributions: one associated with $f_0$, which can be evaluated analytically, and the other from $\delta f$. The simulation particles, along with statistical noise, only contribute to the backreaction from the $\delta f$ part. As a result, when counting the CR backreaction (Equation~\ref{equ::feedback_define}) from $\delta f$, the contribution from each simulation particle is changed by multiplying a weighting factor $w$,
\begin{equation}
    w = 1- \frac{f_0\left(t, \boldsymbol{x}\left(t\right), \boldsymbol{p}\left(t\right)\right)}{f\left(t, \boldsymbol{x}\left(t\right), \boldsymbol{p}  \left(t\right)\right)} \label{equ::weight_delta_f}.
\end{equation}

Traditionally, $f_0$ is often fixed as the initial CR distribution \citep[e.g., ][]{2019ApJ...876...60B, 2022ApJ...928..112B} or as the CR equilibrium distribution after adiabatic evolution \citep[e.g., ][]{2023MNRAS.523.3328S}. With the expanding box, however, the bulk anisotropy level in $f$ keeps changing due to both the $\dot{B}/B$ driving and quasi-linear evolution. This would render the $\delta f$ method less effective in mitigating statistical noise. Therefore, we adaptively adjust $f_0$ towards matching the evolving distribution $f$, so that the statistical noise remains at a relatively low level. 

By design, the adjustment of $f_0$ can be arbitrary. In our case, we initialize the CR particles with an isotropic $\kappa$ distribution \footnote{While analytical and numerical studies of the CR gyro-resonant instability often use a broken power-law distribution \citep{2018MNRAS.476.2779L, 2019ICRC...36..279H, 2019ApJ...882....3H}, we adopt a continuous $\kappa$ distribution for two reasons. Numerically, $f_0$ needs to span the entire momentum range to ensure $w \ll 1$ for the $\delta-f$ method in Equation~\ref{equ::weight_delta_f} to reduce noise \citep{2019ApJ...876...60B}. Physically, it qualitatively mimics more realistic CR distribution with a peak in the spectrum ($\sim$GeV), and allows us to study the scattering for both higher and lower energy particles.}
\begin{equation}
    f_{\text{iso}}\left(\boldsymbol{p}\right) = \frac{\rho_{\text{CR}}}{m \left(\pi  \kappa p_0^2\right)^{1.5}} \frac{\mathcal{G}\left(\kappa + 1\right)}{\mathcal{G}\left(\kappa - 0.5 \right)}\left(1 + \frac{p^2}{\kappa p_0^2}\right)^{-\kappa - 1}, \label{equ::kappa_dist_iso}
\end{equation}
where $\mathcal{G}()$, $\kappa$ and $p_0$ denote the Gamma function, the power index, and the \textbf{initial} peak momentum, respectively. As the background expands/compresses, the CR distribution becomes anisotropic as the bi-$\kappa$ distribution \citep{1991PhFlB...3.1835S} (see Appendix~\ref{sec::app_accuracy}) \footnote{This form is mathematically equivalent with the anisotropic $\kappa$-distribution used in \citet{2023MNRAS.523.3328S} (Equation~(35)), where our $(\xi, p_0)$ \textbf{corresponds to $(\xi^{-1}, p_0/\xi^2)$ in \citet{2023MNRAS.523.3328S}}.
The form adopted here has the advantage of being more conveniently compatible with the expanding box setup. In the absence of MHD waves to isotropize the particles, our expanding box setup leads to $\xi = a(t)$ and $p_0$ remaining constant, whereas $p_0$ in the form used by \citet{2023MNRAS.523.3328S} will always evolve.}:
\begin{align}
	f_{\text{aniso}}\left(t, \boldsymbol{p}\right) =&\frac{\rho_{\text{CR}} \xi^4}{m a^4 \left(\pi \kappa p_0^2\right)^{1.5}} \frac{\mathcal{G}\left(\kappa + 1\right)}{\mathcal{G}\left(\kappa - 0.5 \right)} \notag \\ & \left[1 + \frac{p^2}{\kappa p_0^2}\left(\xi^4 \mu^2  - \xi^2 \mu^2 + \xi^2\right)\right]^{-\kappa - 1}, \label{equ::kappa_dist_aniso}
\end{align}
where $\mu$ and $\xi$ represent the CR pitch angle and the anisotropy parameter. The factor $a^4$ in the denominator accounts for the reduction in lab frame density resulting from the background expansion (see previous subsection). 
One notable fact about this distribution function is that the pitch angle anisotropy is significant for particles whose $p \gtrsim p_0$ but becomes negligible when $p \ll p_0$, as the $\kappa$-distribution yields $f(p\ll p_0, \mu) \sim \text{const}$, which changes little in response to expansion/compression. 

In our simulations, however, the MHD waves resulting from the CRPAI tend to isotropize the CR distribution due to quasi-linear diffusion, reducing the anisotropy at certain energy ranges. Additionally, during scattering with MHD waves, the bulk CR population can lose or gain energy, and $p_0$ may shift. We thus parameterize the above form of the anisotropic $\kappa$ distribution (Equation~\ref{equ::kappa_dist_aniso}) as the ansatz for $f_0$, and dynamically estimate $\xi$ and $p_0$, by taking the generalized method of moments for Equation~\ref{equ::kappa_dist_aniso} after treating $f\left(t, \boldsymbol{p}\right)$ as $f_\text{aniso}$:
\begin{align}
	\xi =& \frac{2}{\pi} \frac{\int \dd^3 \boldsymbol{x} \dd^3 \boldsymbol{p} f p \sqrt{1-\mu^2} }{\int \dd^3 \boldsymbol{x} \dd^3 \boldsymbol{p} f p \left|\mu\right| }, \\
	p_0 =&\frac{\sqrt{\pi \kappa} \left(\kappa - 1\right) \mathcal{G}\left(\kappa - 0.5\right)}{2 \mathcal{G}\left(\kappa + 1\right)} \frac{\int \dd^3 \boldsymbol{x} \dd^3 \boldsymbol{p} f p \sqrt{\xi^4 \mu^2  - \xi^2 \mu^2 + \xi^2}  }{\int \dd^3 \boldsymbol{x} \dd^3 \boldsymbol{p} f}. \label{equ::adapt_delta_f_fit_p_0_full}
\end{align}
Note that estimating $\xi$ only requires the information about $f$, but the estimation on $p_0$ is made convenient when $\xi$ is already known. We also mention that while $\rho_{\text{CR}}$ and $\kappa$ can also be adaptive free parameters, in this work, they are taken as fixed. We expect the CR distribution to be largely uniform in our simulations and thus $\rho_{\rm CR}$ is known from the CR mass conservation in the comoving frame. This is also why the integration above is conducted over space which also serves to substantially reduce noise. We also find changes in the parameter $\kappa$ in our simulations to be minor.

The evaluation of $\xi$ and $p_0$ requires integration over $f$, which is again done by the $\delta f$ method using $f_0$ from the previous time-step, denoted as $f_{0,\text{old}}$. The result is
\begin{align}
	&\xi_{\text{new}} = \frac{2}{\pi } \notag \\ & \times \left( \frac{\sqrt{\pi} p_{0,\text{old}} \mathcal{G}\left(\kappa + 1\right) N_{\text{CR}} }{ 2  \xi_{\text{old}} \sqrt{\kappa}\left(\kappa - 1\right) \mathcal{G}\left(\kappa - 0.5\right)}+\int \dd^3 \boldsymbol{x} \dd^3 \boldsymbol{p} \delta f_{\text{old}} p \sqrt{1-\mu^2}\right)  \notag \\ & / \left(\frac{ p_{0,\text{old}} \mathcal{G}\left(\kappa + 1\right) N_{\text{CR}} }{ \xi_{\text{old}}^2 \sqrt{\pi \kappa}\left(\kappa - 1\right) \mathcal{G}\left(\kappa - 0.5\right)} + \int \dd^3 \boldsymbol{x} \dd^3 \boldsymbol{p} \delta f_{\text{old}} p \left|\mu\right| \right), \label{equ::adapt_delta_f_fit_xi} \\
	&N_{\text{CR}} \equiv \frac{\int \dd^3 \boldsymbol{x} \rho_{\text{CR}}}{m a^4}. \notag
\end{align}
and, we utilize $\xi_{\text{new}}$ to calculate $p_{0,\text{new}}$:
\begin{align}
	&p_{0,\text{new}} = \bigg[\frac{\sqrt{\pi \kappa} \left(\kappa - 1\right) \mathcal{G}\left(\kappa - 0.5\right)}{2 \mathcal{G}\left(\kappa + 1\right)} \notag \\ &  \times \int \dd^3 \boldsymbol{x} \dd^3 \boldsymbol{p} \delta f_{\text{old}} p \sqrt{\xi_{\text{new}}^4 \mu^2  - \xi_{\text{new}}^2 \mu^2 + \xi_{\text{new}}^2} \notag \\& + \frac{p_{0,\text{old}} N_{\text{CR}}}{2}\times \left\{
        \begin{array}{rcl}
			\alpha^2 + \frac{\sinh^{-1}\sqrt{\alpha^2 - 1}}{ \sqrt{\alpha^2 - 1}},\ { \alpha \equiv \frac{\xi_{\text{new}}}{\xi_{\text{old}}} >1}\\
			\alpha^2 + \frac{\sin^{-1}\sqrt{1 - \alpha^2}}{ \sqrt{1 - \alpha^2}},\ { \alpha \leq 1}
        \end{array} \right. \bigg] \notag \\ & \bigg/ \left(\int \dd^3 \boldsymbol{x} \dd^3 \boldsymbol{p} \delta f_{\text{old}} + N_{\text{CR}}\right). \label{equ::adapt_delta_f_fit_p_0}
\end{align}
We have validated the above implementation by observing the adiabatic evolution of $f$ in an expanding box in Appendix~\ref{sec::app_accuracy}. Our implementation for the adaptive $\delta f$ method attains a similar signal-to-noise level in the original $\delta f$ method but with more than four times the number of simulation particles  (Appendix~\ref{sec::app_performance}). The computational cost for adaptively fitting $f_0$ is not negligible. We thus fit once after a fixed time interval $\Delta T_{\text{adapt}}$.

\subsection{Numerical setup}
\label{sec::setup}

Our simulation setup closely follows that of \cite{2019ApJ...876...60B} and the CRPAI simulations of \citet{2023MNRAS.523.3328S}, which we briefly describe here. We initialize an isotropic homogeneous $\kappa$ distribution for CRs in a 1D expanding box (Equation (\ref{equ::kappa_dist_iso})) along the $x-$direction and with a constant collision frequency for ion-neutral damping $\nu_{\text{IN}}$. The background gas (thermal ions) has uniform density $\rho_0$ and uniform magnetic field $B_0$ along $\hat{x}$. The expansion rates along the three directions $(\hat{x},\hat{y},\hat{z})$ are set to $(a^2(t), a(t), a(t))$ as mentioned earlier, with $a(t)=\exp\left(\dot{a}t\right)$. The background gas is initialized with a series of forward and background-traveling, left- and right-polarized Alfv\'en waves with random phases, with initial wave intensity spectrum as $|k|I(k) /  B_0^2 = A^2$, where $k$ is the wavenumber, so that the wave energy is equally distributed in logarithmic $k$ space, and the normalization is such that the total wave energy integrated over $k$ is a fraction $2A^2\ln10$ of the background field energy. For the CRPAI, the most unstable wavenumber is given by $k_0 \sim (\Omega m ) / p_0$ \citep{2023MNRAS.523.3328S}. Our simulation box size $L_x$ is thus chosen to accommodate multiple most unstable wavelengths in the comoving frame of the expanding box.
In the simulations, the numerical units include the initial mass density $\rho_0 = 1$, the initial intensity of the background magnetic field $B_g(t=0) = B_0 = 1$, the Alfv\'en velocity of thermal ions $U_A \equiv B_0 / \sqrt{\rho_0} = 1$, the initial cyclotron frequency $\Omega_0 \equiv (q / (mc)) B_0 = 1$, and the initial ion inertial length $\lambda_i=\Omega_0/U_A = 1$.

To integrate the governing equations (Equation~\ref{equ::mass_conserve} to Equation~\ref{equ::pic}), we employ the two-stage van Leer time integrator \citep{2009NewA...14..139S} in \texttt{ATHENA++} \citep{2020ApJS..249....4S,2023MNRAS.523.3328S}. We use the Roe solver \citep{1981JCoPh..43..357R} with third-order reconstruction to solve the MHD equations, and the Boris pusher \citep{boris1972proceedings} to integrate CR particles. Interpolation of fluid quantities and CR backreaction deposition both follow the triangular-shaped cloud scheme \citep{birdsall2004plasma}. To enhance the numerical accuracy in CR backreaction, we enable the adaptive $\delta f$ method (see Section~\ref{sec::adapt_delta_f}) with $\Delta T_{\text{adapt}} = 500 \Omega_0^{-1}$, and divide the CR population into eight momentum bins \citep[e.g. ][]{2019ApJ...876...60B} which span the momentum range from $p_0/500$ to $500 p_0$ in a logarithmic uniform manner. The periodic boundary condition is applied to all fluid quantities and CR particles, except that the gyro phase of particles is randomized upon box-crossing to facilitate quasi-linear evolution \citep{2019ApJ...876...60B}.

Given the extreme level of scale separation in realistic ISM or similar conditions\footnote{The typical ISM values we adopt here include the magnetic field intensity $5 \mu \text{G}$, \textbf{the number density $\sim 30 \text{cm}^{-3}$, the ionization fraction $\sim 10^{-4}$ (for the cold neutral medium)}, and the cosmic ray number density $\sim 10^{-9}\text{cm}^{-3}$.} (e.g., $U_A/c\sim10^{-4}$, $\rho_{\rm CR}/\rho_0\lesssim10^{-6}$), it is impractical to adopt such values in our simulation parameters. Instead, we choose parameters in the same regime as realistic conditions, and the results should be rescaled for real applications. Our default simulation parameters include: $\mathbb{C} = 200 U_A$, $p_0 = m \mathbb{C}$, $\kappa  = 1.25$ (so that $f(p)\propto p^{-4.5}$ at large $p$), $\rho_{\text{CR}} = 4 \times 10^{-5} \rho_0$, $\nu_{\text{IN}} = 2 \times 10^{-5} \Omega_0$, $\dot{a}= \pm 5 \times 10^{-7} \Omega_0$. Note that our choice of $\dot{a}$ and $\nu_{\rm IN}$ is such that $\dot{a} \ll \nu_{\text{IN}} \ll \omega(k_0) \ll \Omega_0$. This ensures that compression/expansion is adiabatic and that $\nu_{\text{IN}}$ is much lower than the frequency of the Alfv\'en modes in our simulation box \citep{2021ApJ...914....3P}. Moreover, the anticipated CR anisotropy level $\left|\xi^2 - 1\right| \sim \left(\nu_{\text{IN}} / \Omega_0\right) \left(\rho_0 / \rho_{\text{CR}}\right) \left(U_A / c\right)$ is kept low at $\sim10^{-2}$ (see Section~\ref{sec::theory_saturation}). Our simulation box size is chosen to be $L_x = 1.92 \times 10^5 U_A/\Omega_0$, corresponding to about 150 most unstable wavelengths ($\approx2\pi p_0/m\Omega_0$), with a resolution of $\Delta x = 5U_A/\Omega_0$ (about 250 cells per most unstable wavelength). In each cell, we initialize 64 simulation particles per momentum bin. Similar to our earlier works, we adopt initial wave amplitudes with $A=10^{-3}/3$ to seed instability growth. Our simulations typically run for a duration of $4.5\times10^{5} \Omega_0^{-1}$. 

We further conduct a brief parameter survey, where we reduce each of the parameters ($\mathbb{C}$, $\rho_{\text{CR}}$, $\nu_{\text{IN}}$, $\dot{a}$) by half while keeping the other physical parameters fixed to the fiducial values. Additionally, we carry out high-resolution simulations with the fiducial physical parameters but a reduced MHD cell size of $\Delta x = 2.5U_A/\Omega_0$, and using 48 simulation particles per momentum bin per cell.  For the run with $\mathbb{C}$ being half of the fiducial value, the most unstable wavelength is also halved. To achieve equivalent resolution of the most unstable wave, we use the same MHD cell size and number of particles per cell as in the high-resolution run.

\section{Theoretical framework}
\label{sec::theory}

Under our simulation setting, the expanding/compressing box continuously drives CR anisotropy, leading to the development of the CRPAI. The wave growth is countered by the ion-neutral damping, reaching a balance in the steady state. In the meantime, the waves attempt to isotropize the CRs, balancing the driving force. In this section, we estimate the steady state anisotropy level, wave amplitudes, and the resulting CR scattering rates by quasi-linear theory (QLT). The result is to be compared to and calibrated by our simulation results to be presented afterward. \textbf{Eventually, we aim at incorporating the anisotropy level and the CR scattering rate into CR (magneto-)hydrodynamics as a reliable subgrid model in the ISM.}

\subsection{Prediction on CR anisotropy level and effective scattering rate at the saturated state}
\label{sec::theory_saturation}

The saturated state is characterized by the aforementioned two balance relations, which we address in the following two subsections. Altogether, they form our version of QLT to predict the CR anisotropy level and effective scattering rates.

\subsubsection{Balancing wave growth and damping}

First, the waves reach a stable amplitude through the competition between CRPAI and damping mechanisms. We extend the analytical study of CRPAI \citep{2023MNRAS.523.3328S} from the non-relativistic regime to the relativistic regime in the expanding box, 
\begin{equation}
\begin{split}
	\Gamma_{\text{growth}} \left(k\right) \approx& - \frac{\rho_{\text{CR}}}{2\rho_0} \Omega \left[1 \pm \left(\xi^2-1\right) \frac{\Omega_0}{k U_A} \right] Q_2 \left(k\right) \xi^2, \\
    \approx& \mp \frac{\rho_{\text{CR}}}{2\rho_0} \Omega \left(\xi^2-1\right) \frac{\Omega_0}{k U_A} Q_2 \left(k\right) \xi^2\ ,
\end{split} \label{equ::growth_rate}
 \end{equation}
 where
 \begin{align}
	Q_2\left(k\right) &\equiv \frac{\sqrt{\pi}}{\kappa^{1.5}}\frac{\mathcal{G}\left(\kappa + 1\right)}{\mathcal{G}\left(\kappa - 0.5 \right)} \frac{m\Omega_0}{k p_\text{peak}} \left(1 + \frac{m^2\Omega_0^2}{\kappa k^2 p_\text{peak}^2}\right)^{-\kappa},\label{equ::gyro_resonance_q} \\
    p_\text{peak} &\equiv p_0 a^{-4/3}(t).
\end{align}
The linear growth rate $\Gamma_{\text{growth}}$ depends on the wavenumber $k$ and the anisotropy parameter $\xi$, and the $\pm$ sign denotes the polarization direction. The fastest growth is achieved at $k_0 \sim \Omega_0 m /p_\text{peak}$ regardless of the propagation direction. We have also dropped $1$ compared to $(\xi^2-1)\Omega_0/kU_A$ in the bracket of Equation~\ref{equ::growth_rate}. This is because growth is the most prominent near $k\sim k_0$, making $\Omega_0/kU_A\sim \mathbb{C}/U_A\gg1$, thus even a weak level of anisotropy would make the second term in the bracket dominate over $1$. \textbf{Although background expansion slowly changes the cyclotron frequency $\Omega = \Omega_0 / a^2(t)$ and the peak momentum $p_\text{peak}$, due to the timescale separation, the background expansion does not affect the instability growth (particle-wave interaction). The resonance condition adapts with the comoving wavelength, such that $k = (\Omega m / p)a^2 = \Omega_0 m / p$.} It should also be noted that this estimation on growth rate assumes that the CR distribution function has the form Equation~\ref{equ::kappa_dist_aniso}, with a uniform anisotropy level at all momenta characterized by $\left|\xi^2 - 1\right|$, which is not necessarily true as the CR particles evolve. Therefore, this estimate should only be taken as a proxy. 

This growth is to be balanced by the wave damping rate. With our approach of ion-neutral damping \citep{2021ApJ...914....3P}, the damping rate is largely constant for all wavelengths we consider ($\nu_\text{IN} \ll k U_A$ ),
\begin{equation}
	\Gamma_{\text{damp}} \left(k\right) \approx \frac{\nu_{\text{IN}}}{2}.
\end{equation}
Balancing growth and damping $\Gamma_{\text{growth}} = \Gamma_{\text{damp}}$, we can estimate the CR anisotropy level\footnote{Here we neglect the wave damping/amplification induced by the background expansion/compression, as $\left|\dot{a}\right| \ll \Gamma_{\text{damp}} \approx \Gamma_{\text{growth}}$.} as
\begin{equation}
	\left|\xi^2 - 1\right| \left(p\right) \approx \frac{m U_A}{p} \frac{\nu_{\text{IN}}}{\Omega} \frac{\rho_0}{\rho_{\text{CR}}} Q_2^{-1}\left(\frac{\Omega_0 m}{p}\right). \label{equ::aniso_theory}
\end{equation}
which, interestingly, depends only on $\nu_{\rm IN}$ but not $\dot{B}/B$. This is the consequence of linear damping and is similar to the case of CR streaming instability, where the expected streaming speed does not depend on CR pressure gradient under linear damping \citep[e.g.][]{2013MNRAS.434.2209W, 2022ApJ...928..112B}. Here, we further approximate $\xi$ to be an explicit function of $p$ (instead of being constant by definition), by substituting $k$ with $\Omega_0 m/p$ for the resonance condition and omitting the dependence on the particle pitch angle $\mu$. 

Furthermore, when treating the CR population as a single fluid, we can generalize the CR anisotropy in terms of CR pressure $P_{\text{CR}}$,
\begin{equation}
	\frac{P_{\text{CR},\perp}}{P_{\text{CR},\parallel}} \equiv \frac{\int \dd^3 \boldsymbol{p} f p^2 \left(1-\mu^2\right)}{2 \int \dd^3 \boldsymbol{p} f p^2 \mu^2}, \label{equ::aniso_def}
\end{equation}
where $\perp$ and $\parallel$ denote the components perpendicular and parallel to the background magnetic field, respectively. If CRs follow the anisotropic $\kappa$ distribution (Equation~\ref{equ::kappa_dist_aniso}), the relation $\xi^2 = P_{\text{CR},\perp} / P_{\text{CR},\parallel}$ holds, and the single-fluid anisotropy level saturates at
\begin{equation}
    \left|\frac{P_{\text{CR},\perp}}{P_{\text{CR},\parallel}} - 1 \right| \approx  1.78 \frac{U_A}{c} \frac{\nu_{\text{IN}}}{\Omega} \frac{\rho_0}{\rho_{\text{CR}}}.
\end{equation}
Here the factor ``1.78'' is derived from $Q_2^{-1}(\Omega_0 m / p_\text{peak})$, where the CRs at the peak of the distribution are taken to be representative of the entire population, as commonly assumed. 

\textbf{Finally, as a sanity check, we compare the expected CR anisotropy level and $U_A/c$ under realistic parameter regimes. Again consider the cold neutral medium, $\nu / \Omega \sim 10^{-7}$, $\rho_{CR} / \rho_0 \sim 10^{-9}$ (see \citealp{2021ApJ...914....3P}, also see Section~\ref{sec::setup}~\&\ref{sec::discuss}), we immediately see that the anticipated anisotropy level exceeds $U_A/c$ by more than an order of magnitude, thus justifying our simplification in Equation~\ref{equ::growth_rate}.}

\subsubsection{Balancing driving and QLD}
\label{sec::nu_theory}
The second balance is on the CR distribution, where driving on CRs by $\dot{B}/B$ should be balanced by CR scattering, which can be expressed under the quasi-linear theory. Using the Fokker-Plank equation, we can write \citep[e.g.][]{1966ApJ...146..480J, 1969ApJ...156..445K, 1975MNRAS.172..557S, 2002cra..book.....S}
\begin{align}
    \partial_t f &+ \frac{\dot{B}}{2B} \left[p \left(1 + \mu^2 \right) \partial_p f + \mu \left(1-\mu^2\right) \partial_\mu f\right]  \notag \\ &= \partial_\mu \left[\left(D_{\mu \mu} \partial_\mu f + \text{sgn} \left(\mu\right)D_{p \mu} \partial_p f \right) \right] \notag \\  +& \frac{1}{p^2} \partial_p \left[ p^2 \left(\text{sgn} \left(\mu\right) D_{p \mu} \partial_\mu f + D_{pp} \partial_p f \right) \right], \label{equ::FP_equ}
\end{align}
where,
\begin{align}
	D_{\mu \mu} =& \frac{\nu \left(1-\mu^2\right)}{2}, \notag \\
	D_{p \mu} =& \frac{\nu \left(1-\mu^2\right)}{2} \frac{p U_A}{v}, \notag \\
	D_{pp} =& \frac{\nu \left(1-\mu^2\right)}{2} \left(\frac{p U_A}{v}\right)^2 . \label{equ::diff_def}
\end{align}
The term proportional to $\dot{B}/B$ corresponds to the anisotropy driving, $\left(\mathbb{D} \cdot \boldsymbol{p} \right) \cdot \partial f / \partial \boldsymbol{p}$ (see Equation~\ref{equ::pic}). Note that it not only affects the pitch angle but also the total momentum. The two terms on the right-hand side represent quasi-linear diffusion (QLD) in the gas comoving frame, resulting from the pitch angle diffusion and CR momentum diffusion when CRs scatter with the Alfv\'en waves. The diffusion coefficients, $D_{\mu \mu}$, $D_{p \mu}$, and $D_{pp}$, are all characterized by a single parameter, the scattering rate $\nu(p, \mu)$ \citep[e.g.][]{1966ApJ...146..480J, 1969ApJ...156..445K, 1975MNRAS.172..557S, 2002cra..book.....S} which, under quasi-linear theory, is given by
\begin{equation}
	\nu\left(p, \mu\right)= \frac{\pi \Omega}{\gamma} \frac{\Omega m}{p \mu} I\left(k_{\text{res}}=\frac{\Omega m}{p \mu}\right) / B_g^2. \label{equ::scattering_rate_qlt}
\end{equation} 

In the saturated state of the CR population, we anticipate a steady ratio of $P_{\text{CR},\perp}/P_{\text{CR},\parallel}$. To proceed, we notice that at a given $p$, we have
\begin{equation}
 \partial_t \ln{\frac{P_{\text{CR},\perp}}{P_{\text{CR},\parallel}}}=\int_{-1}^{1} \partial_tf\bigg[\frac{p^2 \left(1 - \mu^2\right)}{\int \dd \mu f p^2 \left(1 - \mu^2\right)} -\frac{p^2 \mu^2}{\int \dd \mu f p^2  \mu^2}\bigg]\dd \mu.
\end{equation}
We can then integrate Equation~\ref{equ::FP_equ} over the pitch angle $\mu$ in a similar fashion as the above, to yield 
\begin{equation}
     \partial_t \ln{\frac{P_{\text{CR},\perp}}{P_{\text{CR},\parallel}}} + \text{Driving} =  \text{QLD}. \label{equ::FP_aniso}
\end{equation}
Note that the momentum diffusion terms (the second square bracket on the right-hand side of Equation~\ref{equ::FP_equ}) can be dropped because they are of higher order compared to the pitch angle diffusion by a factor of $U_A/c$. In the steady state, we expect that pitch angle scattering can fully balance the anisotropy driving, and we derive the two terms below.

After integration over $\mu$, the driving term becomes
\begin{align*}
	\text{Driving} = \frac{\dot{B}}{2B}& \bigg( \frac{\int \dd \mu \left[ \left( 1 -\mu^4\right) p \partial_p f - \left(1 -  6 \mu^2 + 5 \mu^4\right) f \right]}{\int \dd \mu \left(1-\mu^2\right) f} \notag \\ & - \frac{\int \dd \mu \left[ \left( \mu^2 +\mu^4\right) p \partial_p f - \left(3 \mu^2 -  5 \mu^4 \right) f \right]}{\int \dd \mu \mu^2 f}\bigg).
\end{align*}
To simplify, we assume the CR distribution is close to isotropic (see Section~\ref{sec::setup}), with $P_{\text{CR},\perp} \approx P_{\text{CR},\parallel}$. Ignoring the dependence of $f$ on $\mu$ as a higher-order term and using the isotropic distribution function, $f(p, \mu) \approx f_{\text{iso}}(p)$, the driving term simplifies to
\begin{align}
	\text{driving} \approx -\frac{\dot{B}}{5B} \frac{\partial \ln  f_{\text{iso}}}{\partial \ln p}\ .\label{equ::driving}
\end{align}

The quasi-linear diffusion (QLD) term on the right-hand side of Equation~\ref{equ::FP_aniso}, after integrating over $\mu$ assuming a nearly isotropic distribution, reads
\begin{align*}
	\text{QLD}= - \frac{9}{2 f_{\text{iso}}} \int \dd \mu \frac{\nu \left(\mu-\mu^3\right)}{2} \left(\partial_\mu f + \text{sgn} \left(\mu\right) \frac{p U_A}{v} \partial_p f \right),
\end{align*}
\textbf{where the last line in Equation~\ref{equ::FP_equ} has been dropped as an higher order term compared to the terms in the second line.}
To proceed, we first assume that the small deviation from isotropy can be described by a distribution function that takes the form $f(p, \mu) = f\left(p \sqrt{\xi^4 \mu^2 - \xi^2 \mu^2 + \xi^2}\right)$ (similar to Equation~\ref{equ::kappa_dist_aniso}), with anisotropy level \textbf{$ U_A/c < \left|\xi^2-1\right| \ll 1$, and therefore,
\begin{align*}
    \partial_\mu f \approx \mu \left(\xi^2 - 1\right) \frac{\partial f_\text{iso}}{\partial \ln{p}}, \quad 
    p \partial_p f \approx \frac{\partial f_\text{iso}}{\partial \ln{p}}.
\end{align*}}
Next, the scattering rate is simplified as an effective rate independent of $\mu$,
\begin{equation}
	\nu_{\text{eff}} \left(p\right)\equiv \frac{\int_{-1}^1 \left(1-\mu^2\right) \nu \left(p, \mu\right) \dd \mu}{\int_{-1}^1 \left(1-\mu^2\right) \dd \mu}, \label{equ::eff_nu_def}
\end{equation}
\textbf{so that it can be directly factored out from the integration by substituting $\nu$ with $\nu_{\text{eff}}$. }Under these assumptions, the contribution from QLD can be written in a form similar to Equation~\ref{equ::driving},
\begin{align}
	\text{QLD}\approx - \nu_{\text{eff}} \left(\frac{3}{5} \left(\xi^2 - 1\right) + \frac{9}{8}\frac{U_A}{c} \right)\frac{\partial \ln  f_{\text{iso}}}{\partial \ln p} .
\end{align}
\textbf{Note that the prefactors (for the factors $3/5$ and $9/8$) may vary depending on the specific derivation method used. However, by expanding $f$ to the lowest order to incorporate the anisotropy, the form of the $\text{QLD}$ will remain as presented above.}

By inserting the wave saturation condition ($\Omega_0 / k \sim p/m$ in Equation~\ref{equ::growth_rate}) into the CR anisotropy balance condition ($\text{driving} = \text{QLD}$), we build up the relation between the effective scattering rate and environmental parameters for each CR momentum bin,
\begin{equation}
	\nu_{\text{eff}}(p) \approx \frac{1}{3} Q_2 \left(\Omega_0 m/p\right) \left| \frac{\dot{B}}{B} \right| \frac{c}{U_A} \frac{\Omega}{\nu_{\text{IN}}} \frac{\rho_{\text{CR}}}{\rho_{0}}. \label{equ::nu_eff_environ}
\end{equation}
\textbf{Here, we again omit the factor $U_A/c$ when comparing to $\left|\xi^2 - 1\right| \ll 1$, in order for the clear scaling relation.}
The momentum-dependent factor $Q_2$ peaks around $0.56$ (for $\kappa = 1.25$, the exact value depends on $\kappa$). In the conventional single-fluid treatment of the CRs, the peak growth rate is often employed when calculating the balance between wave growth and damping, which yields
\begin{equation}
	\nu_{\text{eff}}  \approx 0.187 \left| \frac{\dot{B}}{B} \right| \frac{c}{U_A} \frac{\Omega}{\nu_{\text{IN}}} \frac{\rho_{\text{CR}}}{\rho_{0}}. \label{equ::nu_eff_single}
\end{equation}

Combining Equation~\ref{equ::scattering_rate_qlt} and Equation~\ref{equ::nu_eff_environ}, we can further estimate the saturation level of the CRPAI as:
\begin{align}
   \left(\frac{\delta B}{B_g}\right)^2 \sim &  \frac{k I(k)}{B_g^2} \notag \\ \sim & \frac{\sqrt{1 + \Omega^2 / (kc)^2 }}{3 \pi} Q_2 \left(k\right)  \frac{\left|\dot{B}/ B \right|}{\Omega} \frac{c}{U_A} \frac{\Omega}{\nu_{\text{IN}}} \frac{\rho_{\text{CR}}}{\rho_{0}} \notag \\ \ll & 1 \label{equ::saturate_wave_theory},
\end{align}

The theoretical calculations above have been significantly simplified under several approximations, and are intended to provide an order-of-magnitude estimate on how the saturated state responds to the environment. We anticipate the results to provide the appropriate scaling relations, but the pre-factors can be subject to major uncertainties, which we aim to calibrate through our simulations (Section~\ref{sec::result}).

\subsection{Implication of effective scattering rate in CR (magneto-)hydrodynamics}
\label{sec::theory_crhd}
At macroscopic scales (e.g., in the Galaxy), the cosmic rays are typically modeled as fluid, known as CR (magneto-)hydrodynamics, characterized by the CR energy density $\mathcal{E}_{\text{CR}}$ and the CR energy flux density $\boldsymbol{\mathcal{F}}_{\text{CR}}$, 
\begin{align}
	\mathcal{E}_{\text{CR}} \left(t,\boldsymbol{x}\right) \equiv& \int \dd^3 \boldsymbol{p} f \left(t, \boldsymbol{x}, \boldsymbol{p}\right)\sqrt{c^2+p^2 }c \notag \\ \equiv& \int \dd \ln p \mathcal{E}_{\text{CR}} \left(t,\boldsymbol{x}, p\right), \notag \\
	\boldsymbol{\mathcal{F}}_{\text{CR}} \left(t,\boldsymbol{x}\right) \equiv& \int \dd^3 \boldsymbol{p} f \left(t, \boldsymbol{x}, \boldsymbol{p}\right)\boldsymbol{p}c^2 \notag \\ \equiv& \int \dd \ln p \boldsymbol{\mathcal{F}}_{\text{CR}} \left(t,\boldsymbol{x}, p\right), \label{equ::cr_fluid_def}
\end{align}
where we can either approximate CRs as a single fluid or multiple fluids distributed across different energy (momentum) bins. They are coupled with background gas through wave-particle interaction, described by the scattering rates. 

In our simulation setting with an expanding box, the outcome of the CRPAI in balance with ion-neutral damping only affects $\mathcal{E}_{\text{CR}}$, as $\boldsymbol{\mathcal{F}}_{\text{CR}}=0$ in our setups. By integrating the Fokker-Plank equation (Equation~\ref{equ::FP_equ}) over the CR population, and applying similar techniques as described in Section~\ref{sec::theory_saturation}, the CR fluid energy equation reads
\begin{align}
	\partial_t \mathcal{E}_{\text{CR}} -  \frac{8}{3} \frac{\dot{B}}{B} \mathcal{E}_{\text{CR}} = \frac{U_A}{c} \nu_\text{eff}\mathcal{E}_{\text{CR}} \left(\frac{1}{2}\left(\xi^2 - 1\right) +\frac{4}{3} \frac{U_A}{c} \right).\label{equ::crhd_energy}
\end{align}
The term $8 \dot{B} / \left(3 B\right) \mathcal{E}_{\text{CR}}$ corresponds to the external driving through adiabatic compression/expansion, and the right-hand side corresponds to the quasi-linear diffusion. Note the different pre-factors compared to those in Section~\ref{sec::theory_saturation} due to additional weighting in the moment equations. The above equation can also be naturally re-expressed in the $p$ by $p$ treatment, by substituting $\mathcal{E}_{\text{CR}}\left(t,\boldsymbol{x}\right)$ with $\mathcal{E}_{\text{CR}}\left(t,\boldsymbol{x}, p\right)$. 

In CR (magneto-)hydrodynamics, the anisotropy level $\xi^2 - 1$ and $\nu_\text{eff}$ are expected to be user-specified parameters. Our estimates in Equations~\ref{equ::aniso_theory}~\&~\ref{equ::nu_eff_environ} can be considered as sub-grid prescriptions under the assumption that ion-neutral damping dominates. Under such prescriptions, the QLD effect in the CR energy density equation (the right-hand side of Equation~\ref{equ::crhd_energy}) can be reformulated in terms of environmental parameters as,
\begin{align}
	\partial_t \mathcal{E}_{\text{CR}} -  \frac{8}{3} \frac{\dot{B}}{B} \mathcal{E}_{\text{CR}}\sim - \frac{1}{6} \frac{\dot{B}}{B} \frac{U_A}{c} \mathcal{E}_{\text{CR}}\left(p\right)\ ,
 \label{equ::crhd_energy1}
\end{align}
where we expect the $(\xi^2-1)$ term to dominate over the $U_A/c$ term on the right hand side of Equation~\ref{equ::crhd_energy}. We see that wave scattering on CRs (which is irreversible) always reduces the adiabatic cooling/heating on CR energy density from external driving, but it is only a minor effect of the order $(U_A/c)/16$. Therefore, for the pure CRPAI case, this effect is largely negligible on CR energy density evolution in our simulation setup \citep{2011ApJ...731...35Y, 2020ApJ...890...67Z}.

This work aims to calibrate $\xi^2$ and $\nu_\text{eff}$ in the CR energy equation (Equation~\ref{equ::crhd_energy}) using kinetic simulations. The CR fluid quantities ($P_{\text{CR},\perp}$, $P_{\text{CR},\parallel}$, and $\mathcal{E}_{\text{CR}}$) can be measured by definition. The effective scattering rate of CRSI has been previously measured through the steady-state condition in the CR flux equation \citep{2022ApJ...928..112B}. However, in our simulation, due to the absence of an equilibrium state for the CR energy (Equation~\ref{equ::crhd_energy1}), we cannot perform a similar measurement of $\nu_\text{eff}$ through Equation~\ref{equ::crhd_energy}. Instead, we measure $\nu$ by definition (Section~\ref{sec::result_scatter}) and then numerically integrate the Fokker-Planck equation (Equation~\ref{equ::FP_equ}) for the QLD effect on CR energy density. By comparing Equation~\ref{equ::crhd_energy} with the numerical integration of Equation~\ref{equ::FP_equ}, we ultimately obtain $\nu_\text{eff}$ (Section~\ref{sec::eff_scatter}).

Note that as CR streaming is not present in our problem setup, we do not include the equation for the CR energy flux. In the more general case with CR streaming, the scattering rate will also affect the evolution of the CR energy flux, and $\nu_{\rm eff}$ is likely determined by the combined outcome of the CRSI and CRPAI. This is left for our future work.

\section{Simulation results}
\label{sec::result}

In this section, we illustrate the instability growth and the establishment of steady CR anisotropy in our simulations. The linear growth of CRPAI has been demonstrated in \citet{2023MNRAS.523.3328S}. Here, our primary focus lies in the subsequent evolution leading toward saturation. The final outcome is the effective scattering rate $\nu_{\text{eff}}$ at the saturated state. To compute $\nu_{\text{eff}}$, we assess the CR scattering rate by definition and subsequently average it over the CR pitch angle and energy (momentum magnitude). Given that Equation~\ref{equ::nu_eff_environ} is highly approximate, we would like to reproduce its scaling on environmental parameters and calibrate the coefficients, by comparing different runs at the end of this section.

\subsection{Fiducial run}
\label{sec::fid}
\begin{figure}
	\includegraphics[width=\columnwidth]{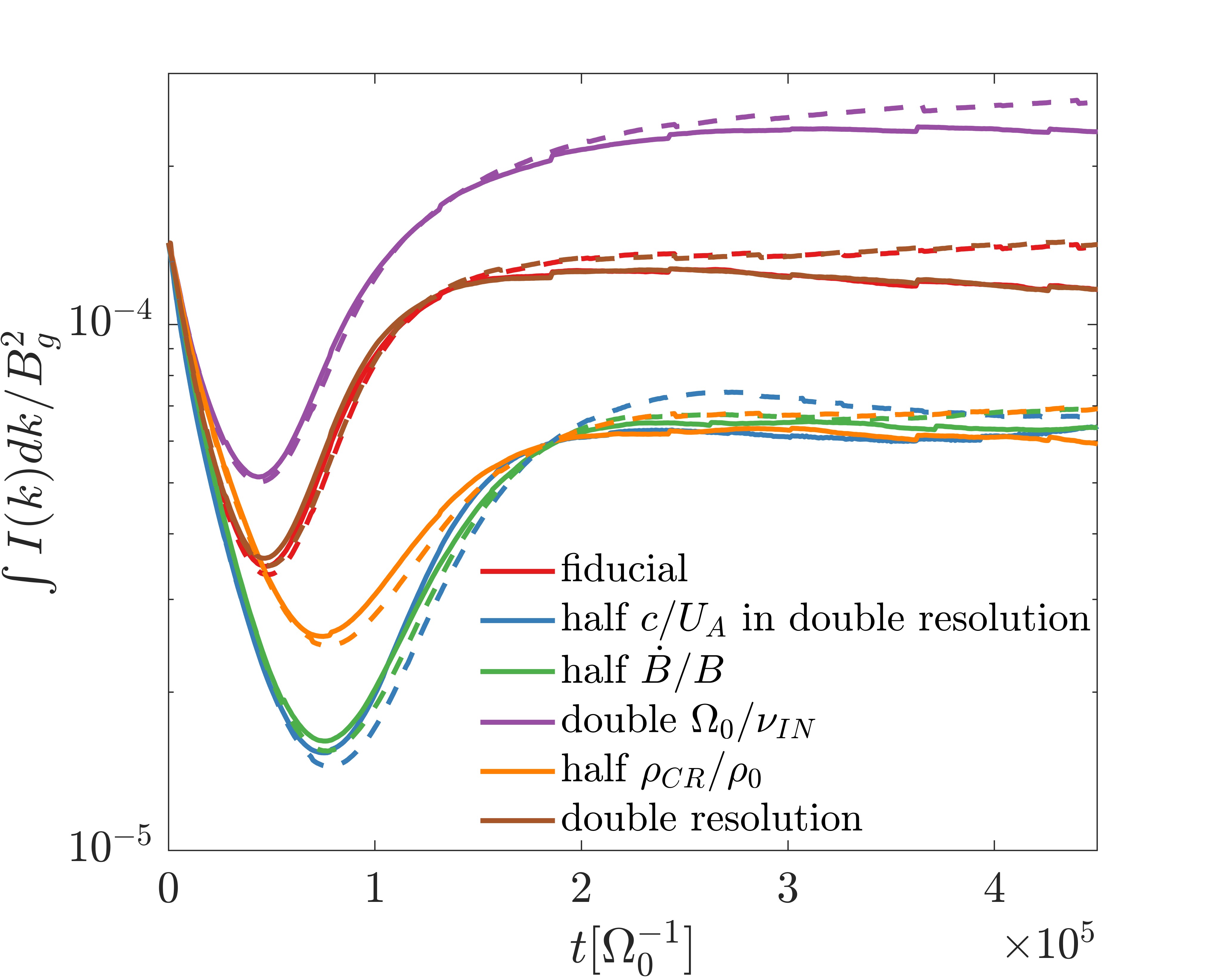}
    \caption{\textbf{Time evolution of the wave energy density around $k_0$, $\int^{5k_0}_{k_0/5} I(k) \dd k B_g^2$. The solid lines represent the compressing box, while the dashed lines denote the expanding box. Line colors distinguish simulations with varying parameters and resolutions. Lines from the runs sharing similar saturated states largely overlap.}}
    \label{fig::wave_hst}
\end{figure}
\begin{figure}
	\centering
     \subfloat[]{
    	\includegraphics[width=\columnwidth]{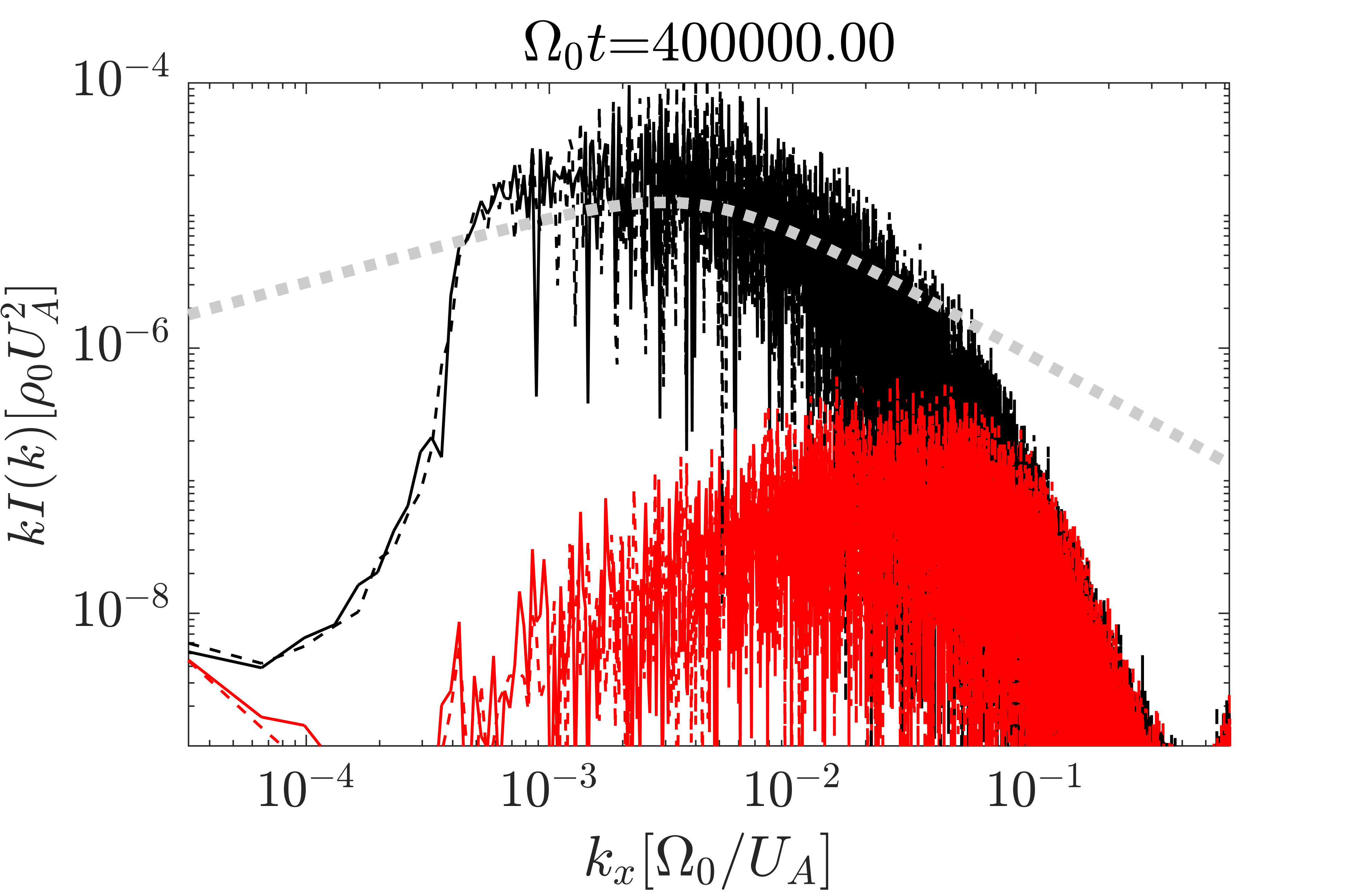}
        \label{fig::expand_fid_wave_spec}
        }
        \\
	\subfloat[]{
    	\includegraphics[width=\columnwidth]{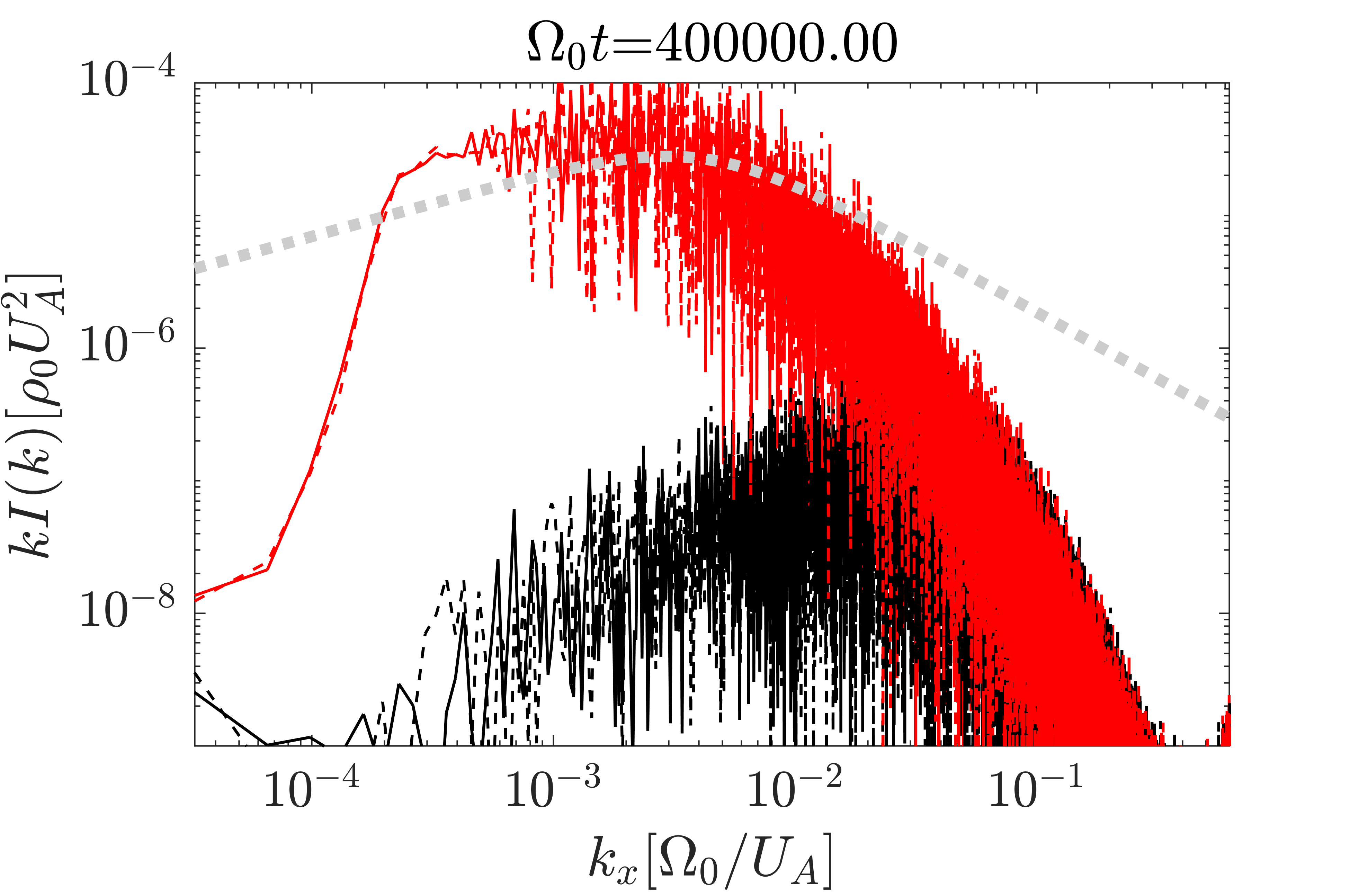}
        \label{fig::compress_fid_wave_spec}
    }
    \caption{Wave intensity spectra, $kI(k)$, during the saturated state of the fiducial runs at $t=4 \times 10^5\Omega_0^{-1}$. The top panel corresponds to the expanding box ($\dot{a} = 5 \times 10^{-7} \Omega_0$), while the bottom panel corresponds to the compressing box ($\dot{a} = -5 \times 10^{-7} \Omega_0$). Solid lines represent forward propagating waves, while dashed lines denote backward propagating ones. The right- and left-handed branches are distinguished by black and red markings. Lines in the same color (polarization) largely overlap. The grey dotted lines refer to the theoretical estimate of the saturated wave intensity (Equation~\ref{equ::saturate_wave_theory}), peaking at the most unstable wave number $k_0\sim 5\times 10^{-3} \Omega_0 / U_A$ for the fiducial runs. The accompanying animation for the wave intensity spectra of the fiducial runs, spanning from time $t=0$ to $4.5 \times 10^5\Omega_0^{-1}$, is available in the online version of the journal.}
	\label{fig::fid_wave_spec}
\end{figure}

We start by showing in Figure~\ref{fig::wave_hst} the time evolution of the wave energy density for all our simulation runs. The initial seed waves are first damped by both the ion-neutral damping and the CRs' gyro-resonance (before $\left|\xi^2 - 1\right| > k U_A / \Omega_0$, the gyro-resonant instability will damp waves (Equation~\ref{equ::growth_rate})). In the meantime, after the CR anisotropy gradually develops owing to $\dot{B}/B$, the CRPAI starts to overcome damping, eventually leading to wave growth. Given our fiducial parameters, the turning point is expected to occur when the CR anisotropy level reaches $\dot{a}t\sim (U_A/\mathbb{C})[1+(\nu_{\rm IN}/\Omega)(\rho_0/\rho_{\rm CR})Q_2^{-1}(\Omega_0 m/p_\text{peak})]$, yielding $t\sim1.5\times10^{4}\Omega_0^{-1}$ for the fastest growing mode in the fiducial case. The actual turn-over when summing over all waves starts later around $t\sim4\times10^{4}\Omega_0^{-1}$ as it takes longer for waves at other wavelength (which grow slower) to catch up. After a few e-folding time, the waves are sufficiently grown to efficiently scatter and isotropize the CR particles. This will reduce the CR anisotropy and hence the wave growth, eventually leading to the saturation of the wave amplitudes, where wave growth due to CRPAI balances the ion-neutral damping.

\subsubsection[]{The wave spectrum}
\begin{figure*}
	\centering
    \subfloat[]{
    	\includegraphics[width=\textwidth]{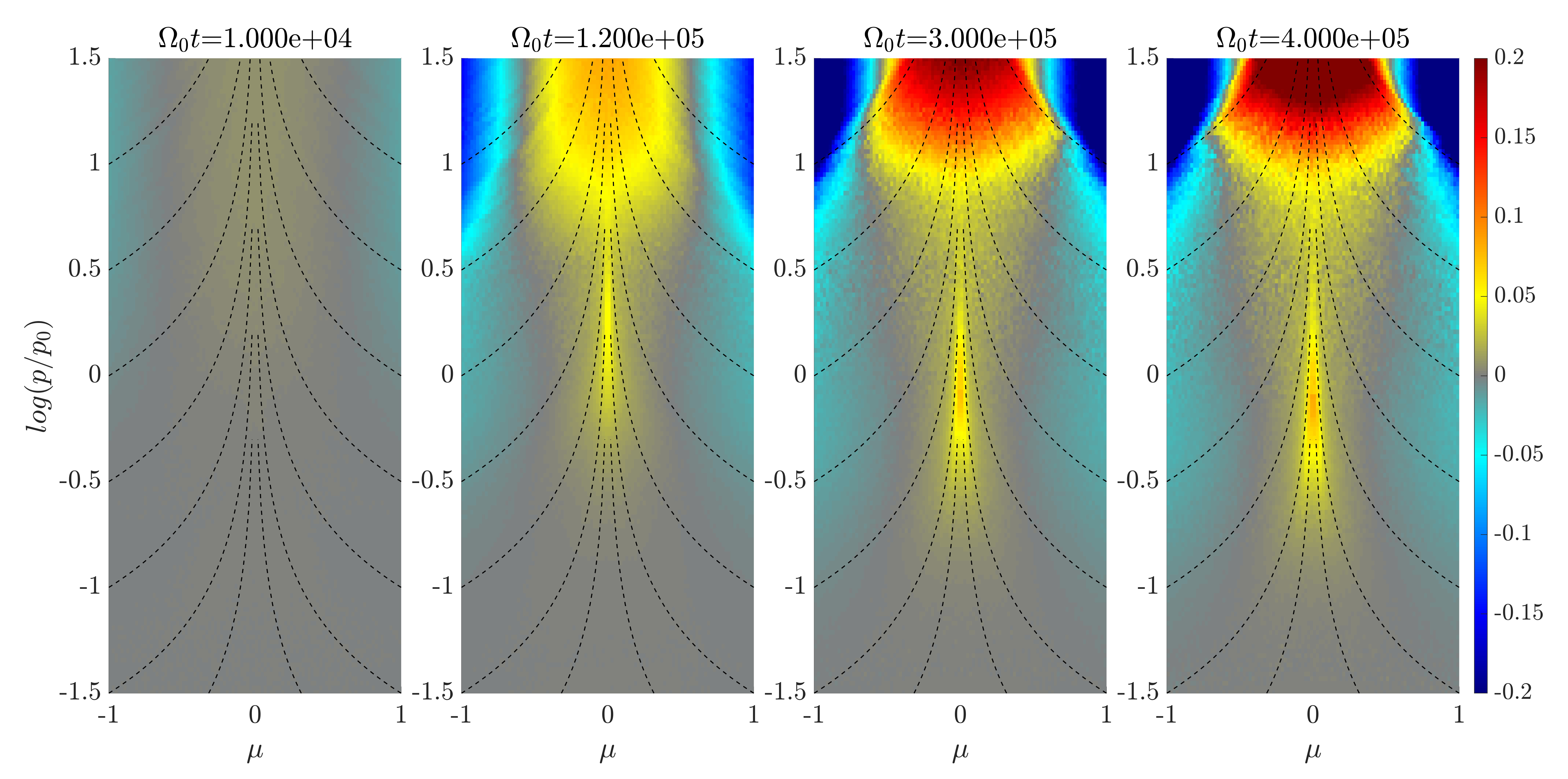}
        \label{fig::expand_fid_par_spec}
    }
    \\
	\subfloat[]{
    	\includegraphics[width=\textwidth]{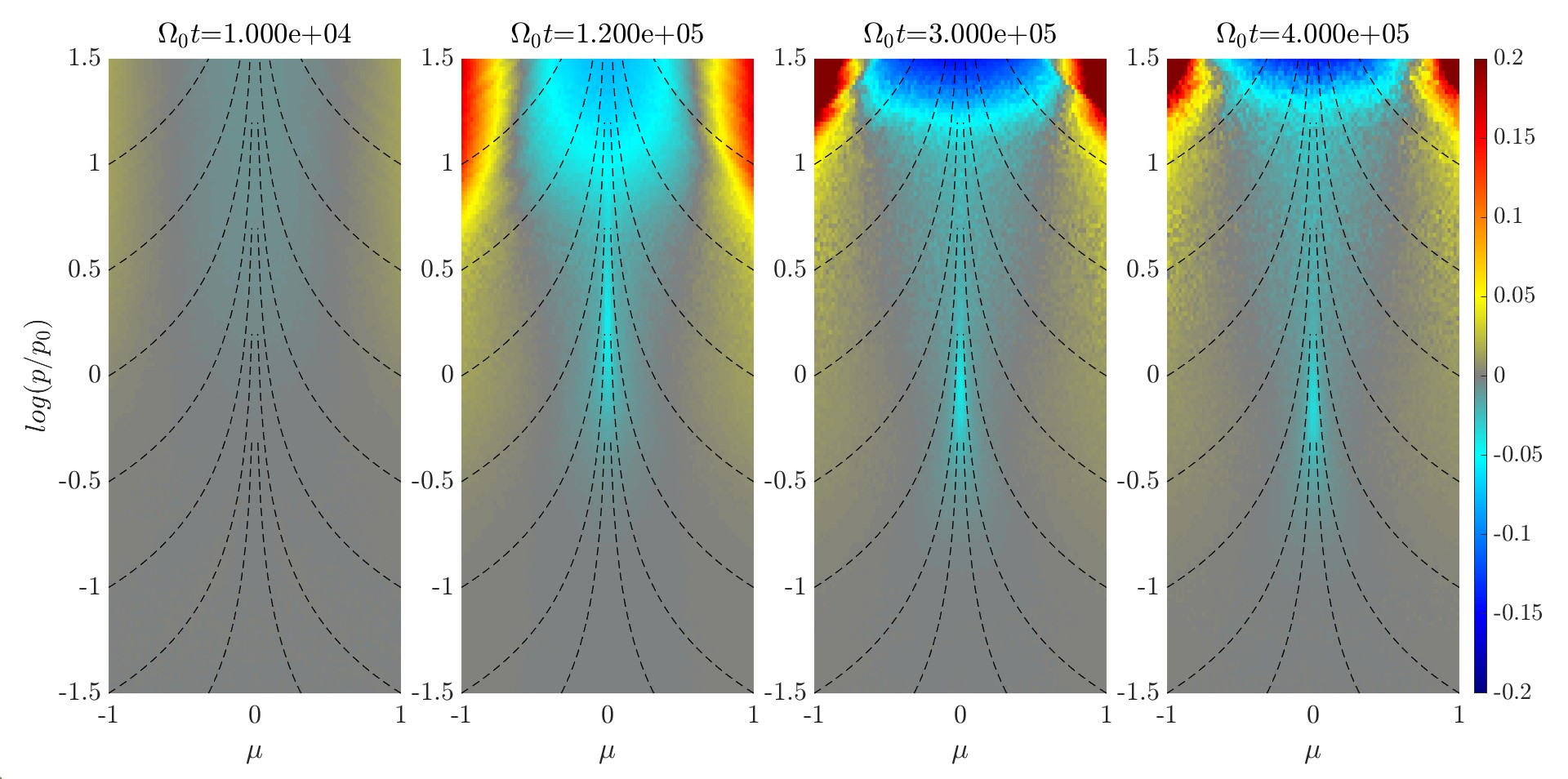}
        \label{fig::compress_fid_par_spec}
    }
    \caption{The CR distribution function $\left(f / \left \langle f \right \rangle_\mu - 1 \right)$ for the fiducial runs in the expanding box (top) and the compressing box (bottom) at various evolution snapshots. Dashed lines represent particle momenta resonating with the same wave characterized by a constant $k=\Omega_0 m / \left(p \mu\right)$. The animation of the CR distribution function of the fiducial runs} spanning from $t=10^3$ to $4.5 \times 10^5\Omega_0^{-1}$ is accessible in the online version of the journal.
	\label{fig::fid_par_spec}
\end{figure*}
\begin{figure}
	\includegraphics[width=\columnwidth]{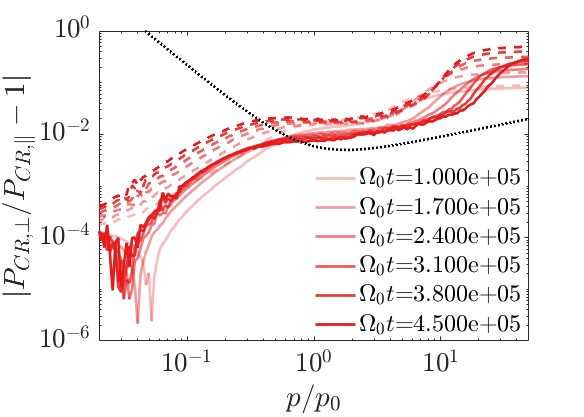}
    \caption{The CR pressure anisotropy level as a function of CR energy (momentum) at different snapshots of the fiducial runs. The solid lines represent the case of the compressing box, while the dashed lines denote the expanding box case. Transparency indicates the simulation run time. The black dotted line corresponds to the theoretical prediction given by Equation~\ref{equ::aniso_theory}.}
    \label{fig::crai_aniso_time}
\end{figure}
To look more closely into the processes, we decompose the waves into the Fourier space, according to polarization and propagation directions \citep{2019ApJ...876...60B}, and plot the wave energy spectra $I(k)$ of the fiducial runs at saturated state in Figure~\ref{fig::fid_wave_spec}. In our setup, the expansion ($\dot{B}/B < 0$) yields an oblate CR distribution ($P_{\text{CR},\perp} / P_{\text{CR},\parallel} > 1$), where the CRPAI amplifies right-handed waves propagating in opposite directions and damps the left-handed counterparts, while the compression ($\dot{B}/B > 0$) leads the waves to saturate at the opposite polarization direction due to the prolate CR distribution ($P_{\text{CR},\perp} / P_{\text{CR},\parallel} < 1$). The dependence of wave intensity on $k$ is consistent with theoretical expectations of the CRPAI. The wave growth rate peaks at the resonant wave number of particles with $p=p_\text{peak}$ given by $k_0 \sim  (\Omega_0 m ) / p_\text{peak}  = 5 \times 10^{-3} \Omega_0 / U_A $ (Equation~\ref{equ::growth_rate}). Consequently, after an initial increase in the CR anisotropy level, the waves around $k_0$ start to grow, and first reach the saturated amplitude (due to QLD, to be discussed next). Waves at other wave numbers require more time to grow and are expected to reach saturation slower with lower amplitudes. 

By the end of the simulations, we observe that waves spanning from $k_0/5$ to $5 k_0$ have approximately reached saturation in the simulations (thus in Figure~\ref{fig::wave_hst} we show the evolution of wave intensity only within this wavenumber range). We also observe from Figure~\ref{fig::fid_wave_spec} that the wave intensity in this spectral range approximately agrees with the estimation from QLT (Equation~\ref{equ::saturate_wave_theory}), \textbf{although with large fluctuations induced by the particle noise in the simulation}.
Note that the final wave intensity slightly differs between the expanding box and the compressing box. We attribute this to the reduction/enhancement of the background magnetic field due to box expansion/compression, which leads to order unity difference in $B_g$ within simulation time, \textbf{while $k I(k)/B_g^2$ largely remains constant (Equation~\ref{equ::saturate_wave_theory} \& Figure~\ref{fig::wave_hst}).}.

\subsubsection[]{Anisotropy in the CR distribution function}

Accompanied by the saturation of wave energy is the asymptotic saturation of the CR distribution $f\left(p, \mu\right)$. Its evolution for the fiducial runs is illustrated in Figure~\ref{fig::fid_par_spec}, which is normalized by the isotropic distribution $\left \langle f \right\rangle_\mu (p)$ averaged over pitch angle. Initially, MHD waves are too weak to scatter the CRs, and hence $f\left(p, \mu\right)$ evolves adiabatically due to expansion/compression approximately following the anisotropic $\kappa$ distribution with $\xi \sim a(t)$ (Equation~\ref{equ::kappa_dist_aniso}): CRs with $p \gtrsim p_0$ concentrate toward $\mu\sim 0$/$\mu = \pm 1$ in expanding/compressing box, while CRs with $p\ll p_0$ in the $\kappa$ distribution do not rapidly develop significant anisotropy (see Section~\ref{sec::adapt_delta_f}). Following the subsequent development of the CRPAI, waves around $k_0$ grow fastest and effectively scatter particles whose momentum is around $p_\text{peak}$. More precisely, scattering is the most effective for particles satisfying $\Omega_0 m /\left(p \mu \right)\sim k_0$ (see dashed lines in Figure~\ref{equ::kappa_dist_aniso}), leading to pitch angle diffusion and hence isotropization. The isotropization gradually propagates towards CRs at lower and higher energies as waves in other wavelengths grow. A similar isotropization process occurs when CRs are initially anisotropic \citep{2018MNRAS.476.2779L}. However, due to the continuous driving of the CR anisotropy, even within the momentum range where scattering is the most effective (around $p=p_\text{peak}=p_0 a^{-4/3}(t)$), the particles are not fully isotropized, leaving a mild anisotropy, reflecting the balance with the box expanding/compressing. The CR anisotropy for particles with $p\in(0.5p_\text{peak}, 5p_\text{peak})$ ultimately reaches a quasi-steady state by the end of the simulations in both the expanding and compressing boxes. Finally, we note that the CR energy distribution keeps evolving following Equations~\ref{equ::crhd_energy}.

We quantify the CR anisotropy level through its pressure anisotropy, $P_{\text{CR}, \perp} / P_{\text{CR}, \parallel} - 1$, based on Equation~\ref{equ::aniso_def}. In the momentum-by-momentum treatment, we present the temporal evolution of CR anisotropy in Figure~\ref{fig::crai_aniso_time}. In the expanding box, the CR pressure perpendicular to the background magnetic field consistently exceeds the parallel component, while the reverse holds in the compressing box. Owing to the inefficient driving of anisotropy towards $p\ll p_0$ (consequence of the $\kappa$ distribution, see footnote 4), we primarily focus on particles with $p\gtrsim0.2p_0$ where driving is effective. Initially, anisotropy levels across all CR momenta increase with time. Subsequently, quasi-linear diffusion (QLD) induced by amplified waves reduces the anisotropy, ultimately maintaining a steady anisotropic profile. The saturated anisotropy level within the momentum range $\left(0.5p_\text{peak}, 5p_\text{peak}\right)$ is broadly consistent with the trend predicted by Equation~\ref{equ::aniso_theory}, albeit higher than the theoretical value by order unity. We attribute this deviation to the omission of pitch angle dependence in deriving Equation~\ref{equ::aniso_theory}, which is the standard approach in QLT but is prone to error, and a similar situation has been discussed in \citet{2022ApJ...928..112B}. Lower-energy CRs in simulations fall below the predicted anisotropy level, which is likely owing to less efficient anisotropy driving. The anisotropy level in the expanding box  is systematically higher than that in the compressing box, \textbf{as a weaker background magnetic field in the expanding box (see Section~\ref{sec::equation}) necessitates a larger anisotropy to sustain the same instability growth rate (Equation~\ref{equ::growth_rate})}.

\subsection{The pitch angle scattering rate and comparison with quasi-linear theory}
\label{sec::result_scatter}

This subsection introduces the methodology to quantify the pitch angle scattering rate $\nu(p, \mu)$ and subsequently compares the obtained measurements with quasi-linear theory (QLT) as a diagnostics.

We quantify the diffusion coefficients by tracing simulation particles \citep{2018MNRAS.476.2779L, 2021ApJ...920..141B}. \textbf{Particles with similar $(p_*, \mu_*)$ at time $t_0$ ($= 4.4 \times 10^5 \Omega_0^{-1}$) are selected into a group (in practice, we select those within $(0.05p_*, 0.005)$ around $(p_*, \mu_*)$), and the pitch-angle and momentum diffusion coefficients are obtained by linearly fitting the growth of group variances over time.}

\textbf{In our setup, beyond QLT, box expansion/compression directly stretches CR momenta while conserving $a^2 p_\parallel$ and $a p_\perp$ \citep[see Appendix of][]{2023MNRAS.523.3328S}. To account for this, we first remove the effect of adiabatic stretching by reconstructing each particle’s momentum $\boldsymbol{p}_k$ into its un-stretched form $\boldsymbol{p}'_k$:}
\begin{align*}
    \boldsymbol{p}'_k (t) &= \left(p'^x_{k}, p'^y_{k}, p'^z_{k}\right) \\& = \left(p^x_{k} e^{2\dot{a} \left(t-t_0\right)}, p^y_{k} e^{\dot{a}\left(t-t_0\right)}, p^z_{k}e^{\dot{a}\left(t-t_0\right)}\right). 
\end{align*}
The pitch angle diffusion coefficient is thus determined by definition,
\begin{equation}
	D_{\mu\mu} \left(\mu_*, p_*\right)= \frac{\partial \text{Var}\left(\boldsymbol{p}' \cdot \boldsymbol{b}_g / \left|\boldsymbol{p}'\right| \right)_{\boldsymbol{p}_*}}{2 \partial t} \label{equ::diffusion_measure}
\end{equation}
\textbf{where $\text{Var}\left(  \right)_{\boldsymbol{p}_*}$ denotes the variance of simulation particles with $\boldsymbol{p} \approx \boldsymbol{p}_*$ at $t=t_0$}. The measurements for $D_{pp}$ follow the same procedure.

\textbf{We set the maximum duration of particle tracking to be $10^4 \Omega_0^{-1}$, sufficient to capture the diffusion behavior for most particles, and the minimum duration as $10^3 \Omega_0^{-1}$. However, as particles undergo QLD, some may change their pitch angles substantially over this period; thus, we set an upper limit of 0.04 for the pitch angle variance and fit over the time interval before reaching this threshold. \footnote{The momentum magnitude evolves more slowly than pitch angle (see Equation~\ref{equ::diff_def}); hence, we do not impose a threshold for its variance.}}

\begin{figure}
	\subfloat[]{
    	\includegraphics[width=0.53\columnwidth]{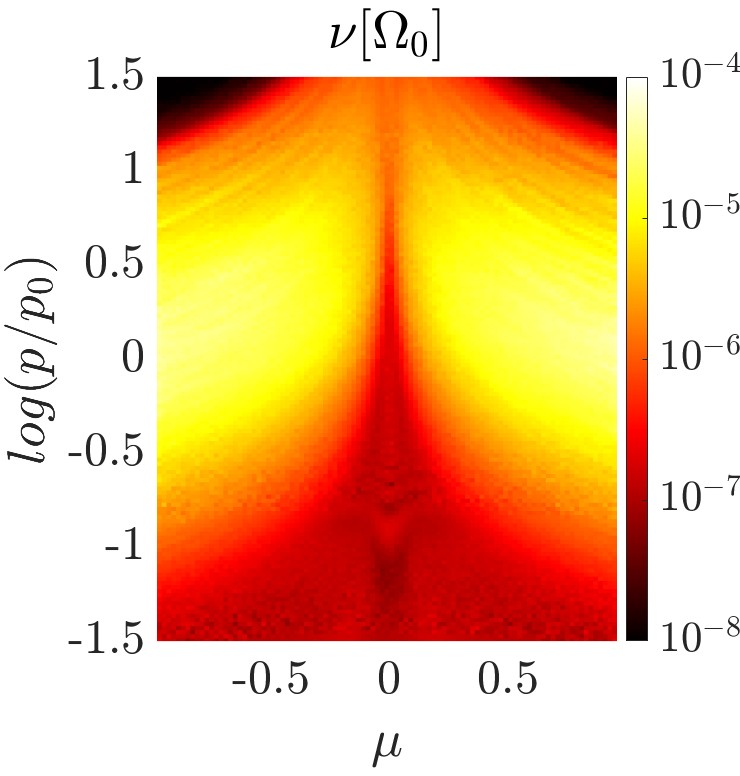}
        \label{fig::scattering_rate_expand_fid}
    }
	\subfloat[]{
    	\includegraphics[width=0.422\columnwidth]{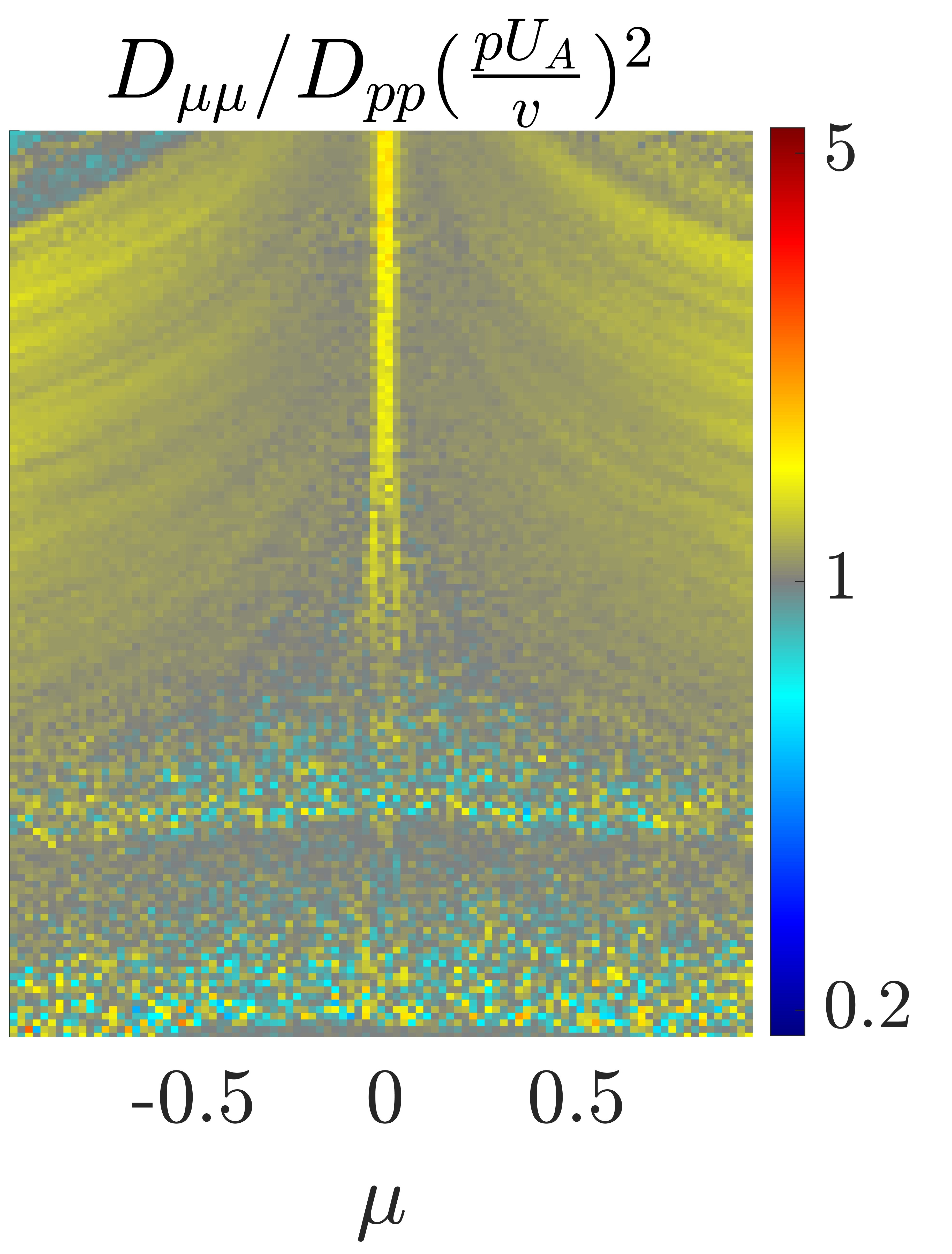}
        \label{fig::D_mumu_over_D_pp_expand_fid}
    }
    \caption{\textbf{\ref{fig::scattering_rate_expand_fid}: The CR scattering rate $\nu$ measured by definition, as a function of CR momentum $p$ and CR pitch angle $\mu$, corresponding to the fiducial run in the expanding box, at the time around $\sim 4.4 \times 10^{5} \Omega_0^{-1}$. \ref{fig::D_mumu_over_D_pp_expand_fid}: The comparison between $D_{\mu \mu}$ and $D_{pp}$ measured by definition. The quasi-linear theory predicts $D_{pp} = D_{\mu \mu} \left(p U_A /v\right)^2$ (see Equation~\ref{equ::diff_def}).}}
\end{figure}
\begin{figure}
	\includegraphics[width=\columnwidth]{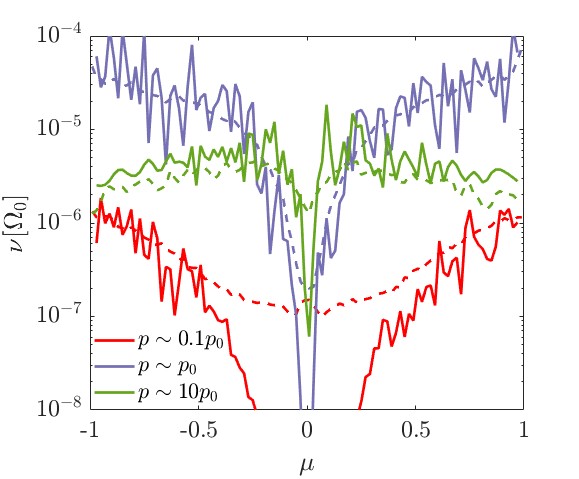}
	\caption{\textbf{The comparison between the pitch angle scattering rate $\nu(p, \mu)$ as a function of $\mu$ for different momenta measured by definition (dashed lines, Equation~\ref{equ::diffusion_measure}) and the one calculated from the wave spectrum (Equation~\ref{equ::scattering_rate_qlt}, solid lines), for the fiducial run in the expanding box at time around $\sim 4.4 \times 10^{5} \Omega_0^{-1}$.}}
    \label{fig::scattering_rate_qlt_expand_fid}
\end{figure}

With the pitch angle diffusion coefficient $D_{\mu\mu}$, the scattering rate, $\nu \left(\mu, p\right)$, is systematically measured as a function of both pitch angle $\mu$ and momentum magnitude $p$. Figure~\ref{fig::scattering_rate_expand_fid} illustrates $\nu \left(\mu, p\right)$ at the saturated state of the fiducial run in the expanding box. The scattering rate exhibits a peak around the momentum $p_0$, corresponding to particles scattering with the most unstable wave at $k_0$, while it diminishes at $\mu = 0$ due to the lack of resonant waves with wave numbers $k \gg k_0$ (see Figure~\ref{fig::expand_fid_wave_spec}). In the context of CR streaming instability, it causes the well-known $90^\circ$ pitch angle crossing problem, where this crossing is necessary to reduce the drift anisotropy \citep{2001ApJ...553..198F, 2019ApJ...882....3H, 2019ApJ...876...60B, CRSI2D}. Given the symmetry with respect to $90^\circ$ pitch angle in the pressure anisotropy problem considered here, this crossing does not affect the isotropization process and hence we do not discuss it further.

The measured value of $\nu$ remains largely consistent regardless of whether it was made through $D_{\mu \mu}$, $D_{p \mu}$, or $D_{pp}$. For example, there exists a pre-factor difference of $\left(p U_A/ v\right)^2$ when calculating from $D_{\mu \mu}$ and $D_{pp}$ (Equation~\ref{equ::diff_def}), which is verified in Figure~\ref{fig::D_mumu_over_D_pp_expand_fid} for the fiducial run in the expanding box. For the most part, the measured ratio $D_{\mu \mu} \left(p U_A /v\right)^2/D_{pp}$ stays around 1 as expected. There is larger numerical noise in the region $p<p_0^{-0.5}$ due to lower wave amplitudes at high-$k$.

Additionally, we compare $\nu$ measured from simulation particles and evaluated through the wave intensity based on QLT (Equation~\ref{equ::scattering_rate_qlt}) in Figure~\ref{fig::scattering_rate_qlt_expand_fid}. \textbf{In accordance with QLT predictions, these two independent measurements show general consistency. Exception is the region near $\mu \sim 0$, likely due to limitations of the measurement method and the influence of nonlinear diffusion. This region corresponds to very low $D_{\mu\mu}$ values, requiring significantly longer time for accurate measurement. Additionally, particles near  $\mu \sim 0$ may be influenced by the processes beyond the scope of QLD (e.g. magnetic mirroring \citep{2001ApJ...553..198F}, resonance broadening \citep{Dupree1966, volk1973, 1981A&A....97..259A} and rotational discontinuities \citep{2021ApJ...914....3P}), which may also relates to the deviation from theoretical predictions in $D_{\mu\mu}/D_{pp}$ (Figure~\ref{fig::D_mumu_over_D_pp_expand_fid}).} In the following sections, we still employ $\nu$ obtained from tracing particles in the subsequent calculations.

\subsection{Anisotropy level and effective scattering rate at the saturated state}
\label{sec::eff_scatter}
The final outcomes of our study is to yield the CR pressure anisotropy level, $P_{\text{CR}, \perp} / P_{\text{CR}, \parallel} - 1$, and the effective scattering rate, $\nu_{\text{eff}}$. They are expected to be given either as a single fluid or on $p$-by-$p$ bases, as a function of environmental parameters ($\dot{B}/B$, $\nu_{\rm IN}$, etc.). We study all our simulation results and compare them with the QLT results derived in Section~\ref{sec::theory}.

\subsubsection[]{Momentum-by-momentum results}

\begin{figure}
	\includegraphics[width=\columnwidth]{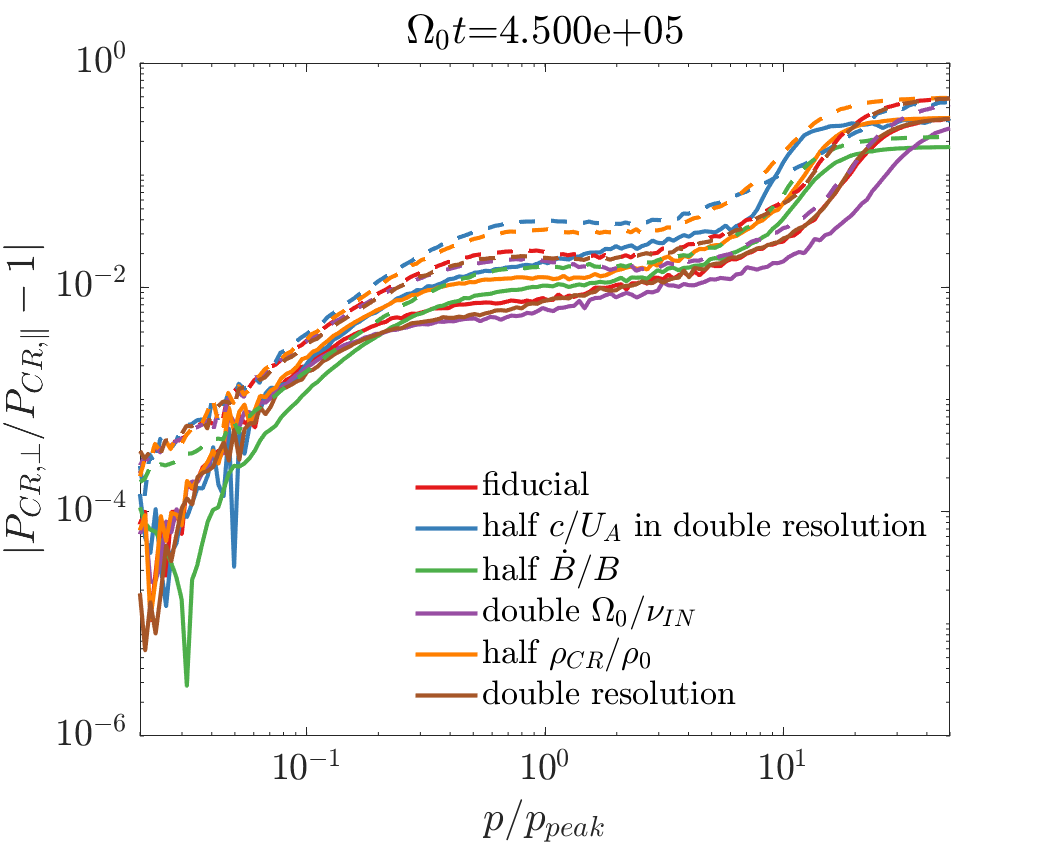}
	\caption{\textbf{The CR pressure anisotropy level, similar to Figure~\ref{fig::crai_aniso_time}, but for different simulation runs at time $4.5 \times 10^5\Omega_0^{-1}$. The solid lines represent the compressing box runs, while the dashed lines denote the expanding box cases. The simulations with varying parameters and resolutions are given in different colors. The particle momentum ($x$-axis) is in units of $p_\text{peak} = p_0 a^{-4/3} (t)$. The animation of the CR pressure anisotropy level spanning from $t=10^3$ to $4.5 \times 10^5\Omega_0^{-1}$ for all runs in different parameters is accessible in the online version of the journal.}}
    \label{fig::crai_aniso_runs}
\end{figure}

The relationship between $P_{\text{CR}, \perp} / P_{\text{CR}, \parallel} - 1$ at the saturated state and environmental parameters is depicted in Figure~\ref{fig::crai_aniso_runs}. The trends of the CR anisotropy level with respect to CR momentum (energy) remain largely similar after varying environmental parameters (see Section~\ref{sec::fid}). The qualitative dependence on magnitude aligns with theoretical expectations (Equation~\ref{equ::aniso_theory}). The expansion/compression rate does not significantly impact CRs at the saturated state (within the momentum range $\left(0.5 p_\text{peak}, 5p_\text{peak}\right)$), as expected with QLT. Reducing $\mathbb{C}/U_A$ or $\rho_{\text{CR}}/\rho_0$, which weakens CRPAI, indeed increases the anisotropy level again consistent with Equation~\ref{equ::aniso_theory}, while we find the anisotropy level is more sensitive to $\mathbb{C}/U_A$ than $\rho_{\text{CR}}/\rho_0$. \textbf{This is owing to a poor scale separation in the simulations when the anisotropy level is comparable to $U_A/c$ (Equation~\ref{equ::growth_rate}).} A weak damping case is counterbalanced by a slow instability growth rate with a smaller anisotropy level, in agreement with results seen in Figure~\ref{fig::crai_aniso_runs}. Within identical environmental parameters, \textbf{stronger expansion yields greater anisotropy than compression, since a weaker background field $B_g$ requires a higher anisotropy to sustain the same growth rate (Equation~\ref{equ::growth_rate})}. We also see that doubling the resolution leads to identical results as illustrated in Figure~\ref{fig::crai_aniso_runs}, thus validating numerical convergence in our simulations.

The effective scattering rate $\nu_{\text{eff}}$, normalized with environmental parameters and as a function of the CR momentum, is illustrated in Figure~\ref{fig::eff_scattering_p_by_p}. The effective scattering rate encapsulates the wave scattering on CR transport, by taking moments of Equation~\ref{equ::FP_equ}. We numerically integrate the Fokker-Planck equation for $\nu_{\text{eff}}$, leveraging the known terms in Equation~\ref{equ::FP_equ} from the simulations. One taking moments method is to retain the dependence of $\nu_{\text{eff}}$ on momentum (energy) $p$, namely the $p$-by-$p$ treatment \citep{2022ApJ...928..112B}. This treatment divides the CR population into distinct momentum (energy) groups, with group $i$ representing a CR fluid composed of CR particles within the momentum range from $p_{i,\text{min}}$ to $p_{i,\text{max}}$. 
Through the definition given by Equation~\ref{equ::aniso_def}~\&~\ref{equ::cr_fluid_def}, we have the energy density and the anisotropy level for the group $i$,
\begin{align*}
    \mathcal{E}_{\text{CR},i} = \int^{p_{i,\text{max}}}_{p_{i,\text{min}}} \dd \ln p\mathcal{E}_{\text{CR}} \left(p \right), \\
    \xi^2_i = \frac{\int^{p_{i,\text{max}}}_{p_{i,\text{min}}} \dd p \int_{-1}^1 \dd \mu f p^4 \left(1-\mu^2\right)}{2 \int^{p_{i,\text{max}}}_{p_{i,\text{min}}} \dd p \int_{-1}^1 \dd \mu f p^4 \mu^2}.
\end{align*}
The QLD effect on $\mathcal{E}_{\text{CR},i}$ can be numerically computed from the right-hand side of Equation~\ref{equ::FP_equ},
\begin{equation*}
\begin{split}
    \text{QLD on group } i = \int^{p_{i,\text{max}}}_{p_{i,\text{min}}} \dd \ln p \int_{-1}^1 \dd \mu p^3 \gamma \mathbb{C}^2\times \\ 
 \frac{\partial}{\partial\ln p} \left[ p^2 f  \times \left( \text{sgn}(\mu)D_{p \mu} \partial_\mu \ln f + D_{pp} \frac{\partial_{\ln p} \ln f}{p} \right) \right], 
\end{split}
\end{equation*}
where both $f$ and the diffusion coefficients are directly measured in the simulations.
By comparing the above numerical integration with the CR energy equation (Equation~\ref{equ::crhd_energy}), we can obtain $\nu_{\text{eff}}\left(p\right)$,
\begin{equation}
	\nu_{\text{eff}}\left(p_i\right) = \frac{\text{QLD on group } i }{\frac{U_A}{\mathbb{C}} \mathcal{E}_{\text{CR},i} \left(\frac{1}{2} \left| \xi^2_i - 1 \right| + \frac{4 U_A}{3 \mathbb{C}}\right)}\ ,
\end{equation}
where $p_i$ should be understood as $\sqrt{p_{i,\text{min}} p_{i,\text{max}}}$ as the mean particle momentum in the $i$th bin.
Here numerical differentiation is applied to the logarithms of $f$ and $p$ to mitigate potential truncation errors. Also, to further minimize errors, we perform one partial integration in computing ``$\text{QLD on group } i$''. 

After normalized by environmental parameters, the pre-factors of $\nu_{\text{eff}}\left(p\right)$ across all simulations, \textbf{both in the expanding boxes and compressing boxes, mostly follow the same curve (Figure~\ref{fig::eff_scattering_p_by_p}) as well as agree with the QLT prediction (Equation~\ref{equ::nu_eff_environ}), which peak around $p_\text{peak}$ where resonant waves are the most pronounced. At $p \ll p_\text{peak}$ and $\mu \sim 0$, numerical dissipation of short-wavelength waves leads to an underestimation of $\nu(\mu, p)$, but this is mitigated at higher resolution, where $\nu_{\text{eff}}$ better approaches QLT values. Compression shifts $p_\text{peak}$ to larger values, worsening the resolution for resonant waves at the same $p/p_\text{peak}$, and consequently reducing $\nu_{\text{eff}}$ in the low-$p$ regime relative to the expanding case. At $p \gg p_\text{peak}$, large fluctuations in $\nu_{\text{eff}}$ arise from non-steady particle anisotropy (Figure~\ref{fig::crai_aniso_time})/ wave amplitudes (Figure~\ref{fig::fid_wave_spec}) and the limited number of resonant wave packets in the simulation domain. Within $(0.5p_\text{peak}, 5p_\text{peak})$, $\nu_{\text{eff}}$ remains close to the theoretical scaling under variations in environmental parameters.} 

\textbf{In both expanding and compressing boxes, the most notable deviation arises in the run with half $\nu_{\rm IN}$, where the prefactor of $\nu_{\text{eff}}$ decreases by approximately 20\% at $p = p_\text{peak}$. Despite this, the integrated saturated wave intensity (Figure~\ref{fig::wave_hst}), dominated by waves resonant with $p_\text{peak}$ (Figure~\ref{fig::fid_wave_spec}), still scales well with environmental parameters. We attribute the discrepancy to limitations in measuring diffusion coefficients (Section~\ref{sec::result_scatter}), where our minimum measuring time is only $10^3 \Omega_0^{-1}$: stronger wave amplitudes enhance particle diffusion, making their momenta significantly deviate within measurement time, leading to an underestimation of $\nu$ for particles near $p = p_\text{peak}$.}

\subsubsection[]{Single-fluid results}
\begin{figure}
	\includegraphics[width=\columnwidth]{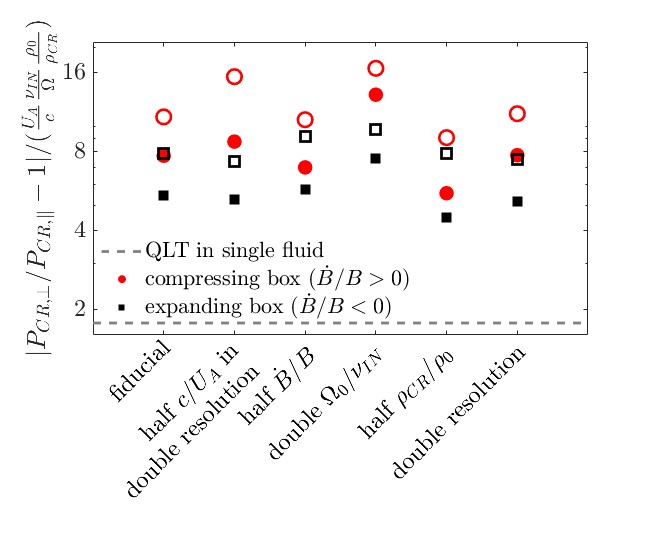}
	\caption{\textbf{The total pressure anisotropy level after normalized by environmental parameters, in the single fluid treatment, \textbf{averaged over  the time $4.4\sim 4.5 \times 10^{5} \Omega_0^{-1}$.} The hollow markers correspond to integrating over majority of the CRs within $0.1 p_\text{peak} < p < 10 p_\text{peak}$, as a CR single fluid, while the filled markers refer to the integration over the CRs which reaches steady pressure anisotropy ($0.5 p_\text{peak} < p < 5 p_\text{peak}$). The red circles correspond to the results in the compressing box and the black squares indicate those in the expanding box. The grey dashed line refers to the theoretical prediction, $|P_{CR,\perp}/P_{CR,\parallel}-1| \sim 1.78\left(U_A/c\right) \left(\nu_\text{IN}/\Omega\right) \left(\rho_0 / \rho_{CR}\right)$ (Equation~\ref{equ::aniso_theory}).}}
    \label{fig::aniso_single}
\end{figure}

\begin{figure}
	\includegraphics[width=\columnwidth]{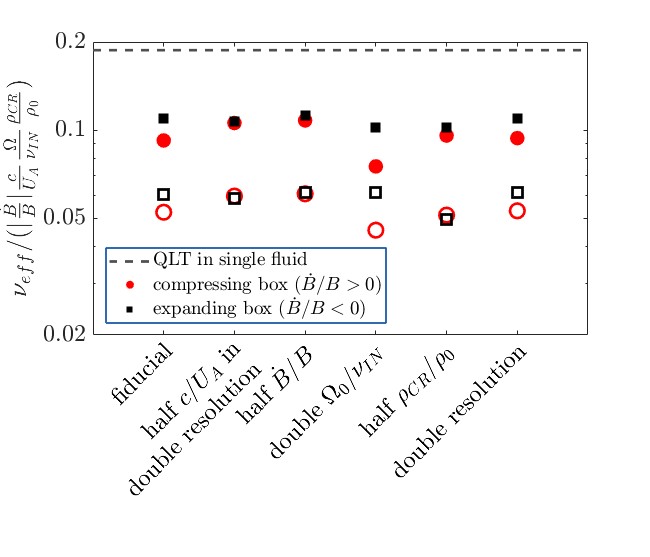}
	\caption{{The effective scattering rate $\nu_{\text{eff}}$ after normalized by environmental parameters, in the single fluid treatment, \textbf{at the time around $\sim 4.4 \times 10^{5} \Omega_0^{-1}$.}. The markers share the same meaning in Figure~\ref{fig::aniso_single}. The grey dashed line refers to the theoretical prediction, $\nu_{\text{eff}} \approx 0.187 \left|\dot{B}/B\right| \left(c/U_A\right) \left(\Omega_0 / \nu_{\text{IN}}\right)$ (Equation~\ref{equ::nu_eff_environ}).}}
    \label{fig::eff_scattering_single}
\end{figure}

More common in the formulation of CR hydrodynamics is to treat the CRs as a single fluid, effectively integrating over the momentum (energy). We determine $\nu_{\text{eff}}$ for the single fluid in the same way, by taking moments for Equation~\ref{equ::FP_equ} and calculating the pressure anisotropy for the whole CR population (Equation~\ref{equ::aniso_def}). By default, we set $p_{\text{max}} = 10 p_\text{peak}$ and $p_{\text{min}} = 0.1p_\text{peak}$, which encompasses the majority of CR particles in the $\kappa$ distribution. However, we also integrate CRs over a short range, $\left(0.5p_\text{peak}, 5p_\text{peak}\right)$, where CR anisotropy reaches a quasi-steady state. We plot the CR anisotropy level and effective scattering normalized by the environmental parameters for all our simulation runs in Figure~\ref{fig::aniso_single}~\&~\ref{fig::eff_scattering_single}.

\begin{figure}
	\includegraphics[width=\columnwidth]{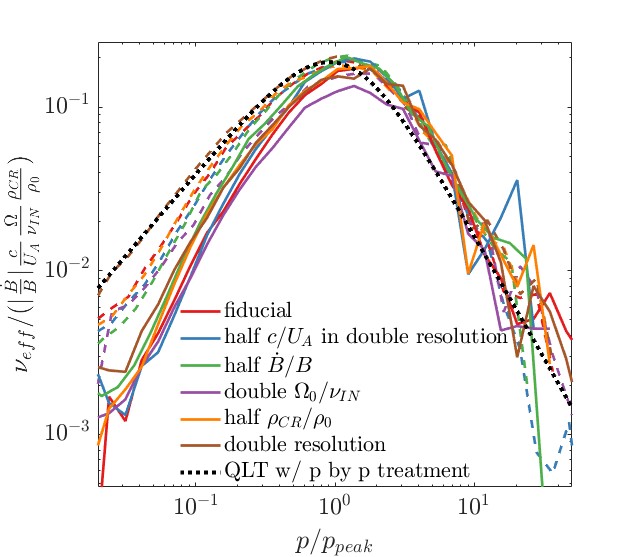}
	\caption{\textbf{The effective scattering rate $\nu_{\text{eff}}$ normalized by environmental parameters, depicted as a function of CR momentum (energy) $p$, \textbf{at the time around $\sim 4.4 \times 10^{5} \Omega_0^{-1}$.} The solid lines represent the compressing box runs, while the dashed lines denote the expanding box cases. The simulations with varying parameters and resolutions are distinguished by line colors, albeit with substantial overlap. The black dotted line illustrates the estimation of quasi-linear theory (QLT) on $\nu_{\text{eff}}$ (Equation~\ref{equ::nu_eff_environ}), in the momentum by momentum treatment.The particle momentum ($x$-axis) is in units of $p_\text{peak} = p_0 a^{-4/3} (t)$.}}
    \label{fig::eff_scattering_p_by_p}
\end{figure}

In Figure~\ref{fig::aniso_single}, we see that the mean CR anisotropy $|P_{CR,\perp}/P_{CR,\parallel}-1|$ for $\left(0.1p_\text{peak}, 10p_\text{peak} \right)$ particles varies between $\sim (8-16)\times\left(U_A/c\right) \left(\nu_\text{IN}/\Omega\right) \left(\rho_0 / \rho_{CR}\right)$, and their corresponding effective scattering rate in Figure~\ref{fig::eff_scattering_single} varies around $0.06 \left|\dot{B}/B\right| \left(c/U_A\right) \left(\Omega / \nu_{\text{IN}}\right) \left(\rho_{\text{CR}}/\rho_0\right)$. 

\textbf{Both the anisotropy level and $\nu_\text{eff}$ deviate from QLT predictions by factors of a few. This discrepancy arises partly because CR pressure anisotropy is dominated by high-energy particles (Figure~\ref{fig::crai_aniso_runs}), as pressure is a second-moment quantity \citep{2013SSRv..175..183L}. Additionally, the standard single-fluid QLT estimate assumes that the entire CR population is represented by particles near $p_\text{peak}$ through the use of $Q_2(\Omega_0 m/p_\text{peak})$. Measurements restricted to $\left(0.5p_\text{peak}, 5p_\text{peak} \right)$ show improved agreement with theoretical expectations.}

\textbf{Although the compressing box cases in Figure~\ref{fig::crai_aniso_runs} exhibit lower absolute anisotropy levels, after normalized by $\Omega/\nu_{\text{IN}}$ which is higher due to stronger background fields, reveals consistently larger normalized anisotropy levels compared to the expanding box. Meanwhile, the most prominent outliers correspond to runs with doubled $\Omega_0/\nu_{\text{IN}}$. We attribute both these deviations to the breakdown of the assumption $\left| P_{\text{CR},\perp} / P_{\text{CR},\parallel} - 1 \right| \gg U_A/c$ in Equation~\ref{equ::growth_rate}, which becomes invalid under weak anisotropy conditions resulting from higher $\Omega/\nu_{\text{IN}}$.}

\textbf{The effective scattering rates, when normalized, show consistency with varying environmental parameters for both expanding and compressing boxes. Applying Chauvenet's criterion \citep{1863mspa.book.....C,1997ieas.book.....T} to the prefactors in Figure~\ref{fig::eff_scattering_single} identifies a single outlier: the half $\nu_\text{IN}$ in the compressing box. Again, this discrepancy is attributed to diffusion coefficient measurement uncertainties (Section~\ref{sec::result_scatter}). The convergence shown in Figure~\ref{fig::eff_scattering_p_by_p} further supports the scaling relation of $\nu_{\text{eff}}$.}

\textbf{Admittedly, the variation in environmental parameters explored in this work is only by a factor of two. This is due to numerical consideration where larger wave amplitude would bring us away from the realistic regime ($\delta B\ll B_0$), while smaller wave amplitude would make simulation time exceedingly long. While it is desirable to extend the simulations further to cover broader range of parameters, the consistency results obtained across all parameter variations and the fact that the prefactor is close to QLT prediction together suggest that our obtained scaling relation is unlikely to be chance coincidence with QLT predictions.}

Finally, the total cooling/heating on the CR energy density, after plugging in $\xi^2$ and $\nu_\text{eff}$ to Equation~\ref{equ::crhd_energy}, yields
\begin{equation*}
    \sim - 0.3 \frac{\dot{B}}{B} \frac{U_A}{c} \mathcal{E}_{\text{CR}},
\end{equation*}
in both expanding boxes and compressing box cases, twice the QLT prediction (Equation~\ref{equ::crhd_energy1}). Therefore, its contribution to irreversible CR heating/cooling raises to about $(1/8)U_A/c$ (instead of $(1/16)U_A/c$) compared to adiabatic heating/cooling, again being very minor to the overall energy.  We find this value is insensitive to variations in environmental parameters, nor to the range of momentum involved (within the range of $(0.1p_\text{peak}, 10p_\text{peak})$ studied here), thanks to the cancellation of such dependencies in $\xi^2$ and $\nu_\text{eff}$.

\section{Discussion: Application of CRPAI}
\label{sec::discuss}

Given the dependency of $\nu_{\text{eff}}$ on environmental parameters, we estimate the value of $\nu_{\text{eff}}$ in realistic systems and compare the importance of CRPAI with that of the Cosmic Ray Streaming Instability (CRSI), which has been widely considered in modeling  CR feedback in the galactic context. The effective scattering rate for the CRSI with ion-neutral damping scales with the environmental parameters as $\sim 0.054 \left(c / L_{\text{gal}}\right) \left(c/U_A\right) \left(\Omega / \nu_{\text{IN}}\right) \left(\rho_{\text{CR}}/\rho_0\right)$ \citep{2022ApJ...928..112B}, where $L_{\text{gal}}$ represents the typical height of the galactic disk scale, also the typical length scale of the CR gradient. Comparing the scaling relations (Equation~\ref{equ::nu_eff_single}), the ratio of $\nu_{\text{eff}}$ between the pure CRPAI (Figure~\ref{fig::eff_scattering_single}) and pure CRSI cases is $5 \left|\dot{B}/B\right| \left(L_{\text{gal}} / c\right)$.

In a general sense, the compression/expansion timescale for the CRPAI can be estimated as the eddy turnover time in turbulence.\footnote{Shear motion can also induce the CR pressure anisotropy, and the local shear timescale in turbulence can be comparable to that of compression/expansion.} Let the largest eddy size in the ISM be $L_{\text{gal}}$, and the eddy turnover timescale for Alfv\'enic turbulence can be estimated as $\left(L_{\text{gal}} / U_A\right) \sim \left|B / \dot{B}\right|$. Therefore, the CR scattering rate resulting from the CRPAI is smaller than that of CRSI by the order of $U_A / c$. A similar conclusion was drawn by \citet{2020ApJ...890...67Z} who focused on CR heating. Setting typical values in the galactic context for the cold neutral medium (CNM) \footnote{Here, we take \textbf{the parameters same as Section~\ref{sec::setup}}, and $L_{\text{gal}} \sim$ 1 kpc.}, one finds $\nu_{\text{eff}}$ for CRSI to be around once per year \citep[e.g., ][]{2019MNRAS.488.3716C, 2020ApJ...890...67Z, 2021MNRAS.501.4184H}, while the effective scattering rate of CRPAI is on the order of $10^{-3}$ per year. Assuming only the CRPAI operates (no CR streaming), we find that the anisotropy level of $\sim \text{GeV}$ CRs is as low as $10^{-3}$ under such parameters.

The CRPAI can be triggered more effectively by smaller eddies. Given an eddy size $l$, one can approximately write $\left|\delta B\right| \sim B_0 \left(l / L_{\text{gal}}\right)^{1/4 \sim 1/3}$ \citep[$1/3$ for the slow magnetosonic mode and $1/4$ for the fast magnetosonic mode, see details in ][]{1995ApJ...438..763G, 2003MNRAS.345..325C}, where $B_0$ represents the mean magnetic field strength. The corresponding variation timescale can be estimated as $l / U_A$, with magnetic field change rate experienced by CRs given by $\left|\dot{B} / B \right| \sim \left|\delta B / B_0 \right| / \left(l / U_A\right) \sim \left(U_A / L_{\text{gal}}\right) \left(l / L_{\text{gal}}\right)^{-3/4 \sim -2/3}$, which increases as the eddy size $l$ decreases. Consider a modest scenario taking $l \sim 1\text{pc}$ while $L_{\text{gal}} \sim 1 \text{kpc}$, the CR scattering rate resulting from the CRPAI may become comparable to that of CRSI in such cases. The transit-time damping in turbulence \citep[e.g. ][]{1976JGR....81.4633F, 1979ApJ...230..373E, 1981A&A....97..259A, 1996ApJ...461..445M, 1998ApJ...492..352S}, which primarily acts on the CR momenta parallel to the background magnetic field, can further enhance CR anisotropy and thereby increase the efficiency of the CRPAI \citep{2007MNRAS.378..245B}. 

Another possible scenario where the CRPAI may actively manifest itself is around the mixing layer of multiphase gas \citep{1990MNRAS.244P..26B, 1994ApJ...420..213K}. Multiphase gas is ubiquitous in astrophysical environments such as the shell of bubbles from stellar explosions \citep[e.g.,][]{2019MNRAS.490.1961E,Lancaster2021,2021ApJ...914...90L,Lancaster2024}, the circumgalactic medium (CGM) \citep[e.g.,][]{Gronke2018,Chen2024}, galactic winds \citep[][]{2022ApJ...924...82F}, and cosmic filaments \citep[e.g.,][]{2020MNRAS.494.2641M}. Mixing layers that separate and mix up different phases at the boundary are dynamic, turbulent and usually rapidly radiating energy away\citep[e.g. ][]{2019MNRAS.487..737J, 2020ApJ...894L..24F,Tan2021}, with the inflow gas experiencing cooling contraction and rapid magnetic field amplification \citep{2023MNRAS.526.4245Z,Das2024}. While the dynamical effect of the CRs in the mixing layers is yet to be understood, one might approximately estimate that the magnetic field changing timescale $|\dot{B}/B|^{-1}$ as experienced by CRs co-moving with the gas can be as short as $\sim 0.01\rm \ Myr$\footnote{Assume $h$ is the thickness of the mixing layer and $v_{\rm in}$ is the inflow velocity of the hot gas. Since the energy radiated by cooling is roughly compensated by the enthalpy flux flowing in, on the order of magnitude we have $h/v_{\rm in}\sim t_{\rm cool}$ where $t_{\rm cool}$ is the shortest cooling time in the mixing layer \citep{Tan2021,Lancaster2024}. With $nt_{\rm cool}=k_B T/\Lambda(T)$ is solely a function of temperature $T$ (where $n$ is the number density and $\Lambda$ is the cooling rate), and assuming solar metallicity, we find $nt_{\rm cool}\sim 0.5\ {\rm Myr}/{\rm cm^3}$ for the cool-warm interface $(T\sim 10^3{\rm K})$ and $nt_{\rm cool}\sim 10^3{\rm yr}/{\rm cm^3}$ for the warm-hot interface $(T\sim 10^5{\rm K})$ (see also \cite{Kim2023}). With a typical $n$ in the cool phase $(T\sim 100{\rm K})$ of the ISM being $\sim 30/{\rm cm^3}$ \citep{Draine2011}, we assume an isobaric condition and estimate $n\sim 3\ (0.03){\rm cm^{-3}}$ in the $T\sim 10^3\ (10^5){\rm K}$ phase, and hence $t_{\rm cool}\sim0.2\ (0.03)\ {\rm Myr}$. For the CGM, we have $n\sim 2\times 10^{-3}{\rm cm^{-3}}$ when $T\sim 10^5{\rm K}$, thus $t_{\rm cool}\sim 0.5\ {\rm Myr}$ \citep{2019MNRAS.487..737J}. The rapid radiative cooling causes compression of the gas by a factor $\sim 100$, and the magnetic fields are thus amplified by a factor $\sim 10-100$. Therefore, we can have a magnetic field changing time scale $|\dot{B}/B|^{-1}\gtrsim h/100v_{\rm in}\sim t_{\rm cool}/100$ as short as $\sim 0.01{\rm Myr}$.}. In comparison to the largest eddy turnover time in ISM, approximately $10 \text{Myr}$, the multiphase gas mixing may efficiently induce CR anisotropy, coupling the CRs through the CRPAI, potentially leading to a $\nu_{\text{eff}}$ up to $\sim$ once per year. A comparison with the CRSI is less straightforward since the level of CR gradient across the mixing layer is unclear, but as significant wave damping is expected in the cold phase, the CR distribution tends to be relatively flat across the phase boundaries \citep{2024arXiv240104169A}.

Another potential applicability of the CRPAI is the CR bottleneck \citep[e.g.,][]{1971ApJ...170..265S, 2017MNRAS.467..646W}, where CRs accumulate in front of a gas density bump (a dip in the CR streaming speed $U_A$), leading to a staircase-like CR pressure structure devoid of gradients on both sides of the bottleneck. Such ``mesoscale" structures could be ubiquitous and play a fundamental role in CR feedback \citep[e.g.,][]{2020ApJ...905...19B,2021ApJ...913..106B,2022MNRAS.513.4464T,2022MNRAS.510..920Q}, yet our exploration and understanding of such structures is still at very early stages. The CR streaming is expected to diminish on the two sides of the bottleneck, and they can be subject to the CRPAI in the presence of $\dot{B}/B$. While it is expected that heating by CRPAI is not necessarily efficient \citep{2020ApJ...905...19B}, it is yet to explore how the additional coupling enabled by CRPAI affects the properties of the CR bottleneck and towards global scales.

\section{Summary and Outlook}
\label{sec::summary}

In this study, we employ kinetic (MHD-PIC) simulations to investigate the saturated state of CRPAI in the presence of ion-neutral damping. The CR anisotropy is continuously driven by the expanding box, which mimics gas expansion and compression in turbulence (and potentially shear motion) and is characterized by the magnetic field changing rate (the expansion rate) $\dot{B}/B$. Meanwhile, the ion-neutral damping is treated in the short-wavelength limit generally applicable in the neutral medium of the ISM, where the ionized gas motion is damped at a rate $\nu_{\text{IN}} / 2$. We achieve the saturated state self-consistently, where the growth rate of the CRPAI (Equation~\ref{equ::growth_rate}) is balanced by the ion-neutral damping rate, while the CR anisotropy is maintained between driving (by expanding box) and quasi-linear diffusion (QLD, Equation~\ref{equ::FP_aniso}). The kinetic simulation at the saturated state enables us to calibrate the scattering rate $\nu_{\text{eff}}$ and the equilibrium pressure anisotropy level $P_{\text{CR}, \perp} / P_{\text{CR}, \parallel} - 1$. By directly comparing simulations with the quasi-linear theory (QLT) predictions (Equations~\ref{equ::aniso_theory} and \ref{equ::nu_eff_environ}) and varying the simulation parameters by a factor of two, we quantitatively calibrate  $\nu_{\text{eff}}$ and $P_{\text{CR}, \perp} / P_{\text{CR}, \parallel} - 1$, and validate their scaling relation on the environmental parameters, including the mass density ratio $\rho_{\text{CR}}/\rho_0$ and the Alfv\'en speed $U_A$,  which are essential for understanding the role of CR feedback at macroscopic (e.g., galactic) scales.

Our main findings are as follows:
\begin{itemize}
    \item We formulate the QLT for CRPAI under the aforementioned balancing relations, which yields the momentum-by-momentum estimates of the CR scattering rates (Equation~\ref{equ::scattering_rate_qlt}) and anisotropy level (Equation~\ref{equ::aniso_theory}) including their scaling with environmental parameters ($\dot B/B$, $\nu_{\rm IN}$, etc.).

    \item We have verified that the CR pitch angle and momentum diffusion coefficients $D_{\mu\mu}$, $D_{pp}$ can be accurately reproduced from the wave intensities based on quasi-linear diffusion, lending support to the general validity of QLT.
    
	\item For those CRs that attain the saturated state (within the momentum range $\left(0.5p_\text{peak}, 5p_\text{peak}\right)$) in our simulations, the CR pressure anisotropy $P_{\text{CR}, \perp} / P_{\text{CR}, \parallel} - 1$ generally follows the momentum dependence predicted by QLT, with high-energy CRs showing greater anisotropy (Figure~\ref{fig::crai_aniso_time}). This pressure anisotropy also scales environmental parameters as expected from QLT (Equation~\ref{equ::aniso_theory}). However, the anisotropy level from the single-fluid treatment is higher than QLT predictions by a factor of $2 \sim 4$ (Figure~\ref{fig::aniso_single}).
    
	\item With the $p$-by-$p$ treatment, the measured CR scattering rates $\nu_{\text{eff}}\left(p\right)$  (Figure~\ref{fig::eff_scattering_p_by_p}) are generally in line with the anticipated trend and the dependence on environmental parameters as predicted by QLT (Equation~\ref{equ::scattering_rate_qlt}). In the single fluid treatment for CRs, the values of $\nu_{\text{eff}}$ are comparable to QLT scaling, though with a factor \textbf{around half} in the normalization (Figure~\ref{fig::eff_scattering_single}).

    \item The contribution of irreversible CR heating/cooling from the CRPAI is about a factor of 2 higher than QLT prediction (Section~\ref{sec::theory_saturation}), $\sim -0.3 \left(\dot{B}/B\right) \left(U_A/c\right) \mathcal{E}_{\text{CR}}$, which represents minor contributions compared to adiabatic heating/cooling.
 
	\item In a broader context, the scattering between CRs and MHD waves resulting from CRPAI is notably weaker compared to that of CRSI, by about a factor of $U_A/c$ on global scales. Nevertheless, we speculate that the CRPAI could play a more dominant role in smaller-scale structures, including the multiphase mixing layers and the CR bottleneck. 
\end{itemize}

Our MHD-PIC simulations on the saturation of the CRPAI, together with that of the CRSI \citep{2022ApJ...928..112B}, provide valuable insights into the microphysics about CR feedback from the first principles. The ultimate goal of such studies is to provide calibrated CR scattering rates from the CR gyro-resonant instabilities that serve as subgrid physics for CR (magneto-)hydrodynamic simulations at macroscopic scales. So far, we have studied the pure CRSI case driven by imposed CR pressure gradient, and the pure CRPAI case driven by $\dot{B}/B$. The methodology laid out in these two works paves the way for us to further explore the most general case, where both a CR gradient and $\dot{B}/B$ are present, which may lead to marked differences from the cases studied so far. In the meantime, future work should explore multi-dimensional effects \citep{CRSI2D}, as well as the case with other wave damping mechanisms, especially the non-linear Landau damping that dominates in the more volume-filling warm medium \citep[e.g.,][]{2021ApJ...922...11A}. Finally, with the adaptive $\delta f$ method, it becomes possible to explore ``mesoscale" phenomena such as the CR bottleneck \citep[e.g.,][]{1971ApJ...170..265S, 2017MNRAS.467..646W, 2022MNRAS.513.4464T, 2022MNRAS.510..920Q} fully kinetically, which would be a major effort to directly link CR feedback from microscopic to macroscopic scales.

\begin{acknowledgments}
We thank Eve Ostriker, Eliot Quataert, Ellen Zweibel, and Alexandre Lazarian for useful discussions, particularly during the Aspen Center for Physics program on ‘Cosmic Ray Feedback in Galaxies and Galaxy Clusters’, as well as the anonymous referee for a constructive report. This work is supported by the National Science Foundation of China under grant No. 12325304, 12342501, by Multimessenger Plasma Physics Center (MPPC, NSF grant PHY-2206607), and in part by the National Science Foundation under Grant No. NSF PHY-2210452. Numerical simulations are conducted in the Orion clusters at the Department of Astronomy, Tsinghua University, and TianHe-1 (A) at the National Supercomputer Center in Tianjin, China.
\end{acknowledgments}

\bibliography{example}

\begin{thebibliography}{}
\expandafter\ifx\csname natexlab\endcsname\relax\def\natexlab#1{#1}\fi
\providecommand{\url}[1]{\href{#1}{#1}}
\providecommand{\dodoi}[1]{doi:~\href{http://doi.org/#1}{\nolinkurl{#1}}}
\providecommand{\doeprint}[1]{\href{http://ascl.net/#1}{\nolinkurl{http://ascl.net/#1}}}
\providecommand{\doarXiv}[1]{\href{https://arxiv.org/abs/#1}{\nolinkurl{https://arxiv.org/abs/#1}}}

\bibitem[{{Achterberg}(1981)}]{1981A&A....97..259A}
{Achterberg}, A. 1981, \aap, 97, 259

\bibitem[{{Adkins} \& {Schekochihin}(2018)}]{2018JPlPh..84a9007A}
{Adkins}, T., \& {Schekochihin}, A.~A. 2018, Journal of Plasma Physics, 84, 905840107, \dodoi{10.1017/S0022377818000089}

\bibitem[{{Adriani} {et~al.}(2011){Adriani}, {Barbarino}, {Bazilevskaya}, {Bellotti}, {Boezio}, {Bogomolov}, {Bonechi}, {Bongi}, {Bonvicini}, {Borisov}, {Bottai}, {Bruno}, {Cafagna}, {Campana}, {Carbone}, {Carlson}, {Casolino}, {Castellini}, {Consiglio}, {De Pascale}, {De Santis}, {De Simone}, {Di Felice}, {Galper}, {Gillard}, {Grishantseva}, {Jerse}, {Karelin}, {Koldashov}, {Krutkov}, {Kvashnin}, {Leonov}, {Malakhov}, {Malvezzi}, {Marcelli}, {Mayorov}, {Menn}, {Mikhailov}, {Mocchiutti}, {Monaco}, {Mori}, {Nikonov}, {Osteria}, {Palma}, {Papini}, {Pearce}, {Picozza}, {Pizzolotto}, {Ricci}, {Ricciarini}, {Rossetto}, {Sarkar}, {Simon}, {Sparvoli}, {Spillantini}, {Stozhkov}, {Vacchi}, {Vannuccini}, {Vasilyev}, {Voronov}, {Yurkin}, {Wu}, {Zampa}, {Zampa}, \& {Zverev}}]{2011Sci...332...69A}
{Adriani}, O., {Barbarino}, G.~C., {Bazilevskaya}, G.~A., {et~al.} 2011, Science, 332, 69, \dodoi{10.1126/science.1199172}

\bibitem[{{Adriani} {et~al.}(2013){Adriani}, {Barbarino}, {Bazilevskaya}, {Bellotti}, {Boezio}, {Bogomolov}, {Bongi}, {Bonvicini}, {Borisov}, {Bottai}, {Bruno}, {Cafagna}, {Campana}, {Carbone}, {Carlson}, {Casolino}, {Castellini}, {De Pascale}, {De Santis}, {De Simone}, {Di Felice}, {Formato}, {Galper}, {Grishantseva}, {Karelin}, {Koldashov}, {Koldobskiy}, {Krutkov}, {Kvashnin}, {Leonov}, {Malakhov}, {Marcelli}, {Mayorov}, {Menn}, {Mikhailov}, {Mocchiutti}, {Monaco}, {Mori}, {Nikonov}, {Osteria}, {Palma}, {Papini}, {Pearce}, {Picozza}, {Pizzolotto}, {Ricci}, {Ricciarini}, {Rossetto}, {Sarkar}, {Simon}, {Sparvoli}, {Spillantini}, {Stozhkov}, {Vacchi}, {Vannuccini}, {Vasilyev}, {Voronov}, {Yurkin}, {Wu}, {Zampa}, {Zampa}, {Zverev}, {Potgieter}, \& {Vos}}]{2013ApJ...765...91A}
---. 2013, \apj, 765, 91, \dodoi{10.1088/0004-637X/765/2/91}

\bibitem[{{Armillotta} {et~al.}(2021){Armillotta}, {Ostriker}, \& {Jiang}}]{2021ApJ...922...11A}
{Armillotta}, L., {Ostriker}, E.~C., \& {Jiang}, Y.-F. 2021, \apj, 922, 11, \dodoi{10.3847/1538-4357/ac1db2}

\bibitem[{{Armillotta} {et~al.}(2022){Armillotta}, {Ostriker}, \& {Jiang}}]{2022ApJ...929..170A}
---. 2022, \apj, 929, 170, \dodoi{10.3847/1538-4357/ac5fa9}

\bibitem[{{Armillotta} {et~al.}(2024){Armillotta}, {Ostriker}, {Kim}, \& {Jiang}}]{2024arXiv240104169A}
{Armillotta}, L., {Ostriker}, E.~C., {Kim}, C.-G., \& {Jiang}, Y.-F. 2024, arXiv e-prints, arXiv:2401.04169, \dodoi{10.48550/arXiv.2401.04169}

\bibitem[{{Bai}(2022)}]{2022ApJ...928..112B}
{Bai}, X.-N. 2022, \apj, 928, 112, \dodoi{10.3847/1538-4357/ac56e1}

\bibitem[{{Bai} {et~al.}(2015){Bai}, {Caprioli}, {Sironi}, \& {Spitkovsky}}]{2015ApJ...809...55B}
{Bai}, X.-N., {Caprioli}, D., {Sironi}, L., \& {Spitkovsky}, A. 2015, \apj, 809, 55, \dodoi{10.1088/0004-637X/809/1/55}

\bibitem[{{Bai} {et~al.}(2019){Bai}, {Ostriker}, {Plotnikov}, \& {Stone}}]{2019ApJ...876...60B}
{Bai}, X.-N., {Ostriker}, E.~C., {Plotnikov}, I., \& {Stone}, J.~M. 2019, \apj, 876, 60, \dodoi{10.3847/1538-4357/ab1648}

\bibitem[{{Bambic} {et~al.}(2021){Bambic}, {Bai}, \& {Ostriker}}]{2021ApJ...920..141B}
{Bambic}, C.~J., {Bai}, X.-N., \& {Ostriker}, E.~C. 2021, \apj, 920, 141, \dodoi{10.3847/1538-4357/ac0ce7}

\bibitem[{{Begelman} \& {Fabian}(1990)}]{1990MNRAS.244P..26B}
{Begelman}, M.~C., \& {Fabian}, A.~C. 1990, \mnras, 244, 26P

\bibitem[{{Berezinskii} {et~al.}(1990){Berezinskii}, {Bulanov}, {Dogiel}, \& {Ptuskin}}]{1990acr..book.....B}
{Berezinskii}, V.~S., {Bulanov}, S.~V., {Dogiel}, V.~A., \& {Ptuskin}, V.~S. 1990, {Astrophysics of cosmic rays} (North-Holland)

\bibitem[{Birdsall \& Langdon(2004)}]{birdsall2004plasma}
Birdsall, C.~K., \& Langdon, A.~B. 2004, Plasma physics via computer simulation (CRC press)

\bibitem[{{Blasi}(2013)}]{2013A&ARv..21...70B}
{Blasi}, P. 2013, \aapr, 21, 70, \dodoi{10.1007/s00159-013-0070-7}

\bibitem[{{Blasi} {et~al.}(2012){Blasi}, {Amato}, \& {Serpico}}]{2012PhRvL.109f1101B}
{Blasi}, P., {Amato}, E., \& {Serpico}, P.~D. 2012, \prl, 109, 061101, \dodoi{10.1103/PhysRevLett.109.061101}

\bibitem[{Boris {et~al.}(1972)Boris, Shanny, of~Naval~Research, \& Laboratory}]{boris1972proceedings}
Boris, J., Shanny, R., of~Naval~Research, U. S.~O., \& Laboratory, N.~R. 1972, Proceedings: Fourth Conference on Numerical Simulation of Plasmas, November 2, 3, 1970 (Naval Research Laboratory).
\newblock \url{https://books.google.com.hk/books?id=zqxSAQAACAAJ}

\bibitem[{{Bott} {et~al.}(2021){Bott}, {Arzamasskiy}, {Kunz}, {Quataert}, \& {Squire}}]{2021ApJ...922L..35B}
{Bott}, A.~F.~A., {Arzamasskiy}, L., {Kunz}, M.~W., {Quataert}, E., \& {Squire}, J. 2021, \apjl, 922, L35, \dodoi{10.3847/2041-8213/ac37c2}

\bibitem[{{Br{\"u}ggen} \& {Scannapieco}(2020)}]{2020ApJ...905...19B}
{Br{\"u}ggen}, M., \& {Scannapieco}, E. 2020, \apj, 905, 19, \dodoi{10.3847/1538-4357/abc00f}

\bibitem[{{Brunetti} \& {Lazarian}(2007)}]{2007MNRAS.378..245B}
{Brunetti}, G., \& {Lazarian}, A. 2007, \mnras, 378, 245, \dodoi{10.1111/j.1365-2966.2007.11771.x}

\bibitem[{{Bustard} \& {Zweibel}(2021)}]{2021ApJ...913..106B}
{Bustard}, C., \& {Zweibel}, E.~G. 2021, \apj, 913, 106, \dodoi{10.3847/1538-4357/abf64c}

\bibitem[{{Butsky} \& {Quinn}(2018)}]{2018ApJ...868..108B}
{Butsky}, I.~S., \& {Quinn}, T.~R. 2018, \apj, 868, 108, \dodoi{10.3847/1538-4357/aaeac2}

\bibitem[{{Chan} {et~al.}(2019){Chan}, {Kere{\v{s}}}, {Hopkins}, {Quataert}, {Su}, {Hayward}, \& {Faucher-Gigu{\`e}re}}]{2019MNRAS.488.3716C}
{Chan}, T.~K., {Kere{\v{s}}}, D., {Hopkins}, P.~F., {et~al.} 2019, \mnras, 488, 3716, \dodoi{10.1093/mnras/stz1895}

\bibitem[{{Chauvenet}(1863)}]{1863mspa.book.....C}
{Chauvenet}, W. 1863, {A manual of spherical and practical astronomy}

\bibitem[{{Chen} \& {Oh}(2024)}]{Chen2024}
{Chen}, Z., \& {Oh}, S.~P. 2024, \mnras, 530, 4032, \dodoi{10.1093/mnras/stae1113}

\bibitem[{{Cho} \& {Lazarian}(2003)}]{2003MNRAS.345..325C}
{Cho}, J., \& {Lazarian}, A. 2003, \mnras, 345, 325, \dodoi{10.1046/j.1365-8711.2003.06941.x}

\bibitem[{{Consolandi}(2014)}]{2014arXiv1402.0467C}
{Consolandi}, C. 2014, arXiv e-prints, arXiv:1402.0467, \dodoi{10.48550/arXiv.1402.0467}

\bibitem[{{Das} \& {Gronke}(2024)}]{Das2024}
{Das}, H.~K., \& {Gronke}, M. 2024, \mnras, 527, 991, \dodoi{10.1093/mnras/stad3125}

\bibitem[{{Dashyan} \& {Dubois}(2020)}]{2020A&A...638A.123D}
{Dashyan}, G., \& {Dubois}, Y. 2020, \aap, 638, A123, \dodoi{10.1051/0004-6361/201936339}

\bibitem[{{Denton} \& {Kotschenreuther}(1995)}]{1995JCoPh.119..283D}
{Denton}, R.~E., \& {Kotschenreuther}, M. 1995, Journal of Computational Physics, 119, 283, \dodoi{10.1006/jcph.1995.1136}

\bibitem[{{Dorfi} \& {Breitschwerdt}(2012)}]{2012A&A...540A..77D}
{Dorfi}, E.~A., \& {Breitschwerdt}, D. 2012, \aap, 540, A77, \dodoi{10.1051/0004-6361/201118082}

\bibitem[{{Draine}(2011)}]{Draine2011}
{Draine}, B.~T. 2011, {Physics of the Interstellar and Intergalactic Medium}, Vol.~19 (Princeton University Press)

\bibitem[{{Dubois} {et~al.}(2019){Dubois}, {Commer{\c{c}}on}, {Marcowith}, \& {Brahimi}}]{2019A&A...631A.121D}
{Dubois}, Y., {Commer{\c{c}}on}, B., {Marcowith}, A., \& {Brahimi}, L. 2019, \aap, 631, A121, \dodoi{10.1051/0004-6361/201936275}

\bibitem[{Dupree(1966)}]{Dupree1966}
Dupree, T.~H. 1966, The Physics of Fluids, 9, 1773, \dodoi{10.1063/1.1761932}

\bibitem[{{Eilek}(1979)}]{1979ApJ...230..373E}
{Eilek}, J.~A. 1979, \apj, 230, 373, \dodoi{10.1086/157093}

\bibitem[{{El-Badry} {et~al.}(2019){El-Badry}, {Ostriker}, {Kim}, {Quataert}, \& {Weisz}}]{2019MNRAS.490.1961E}
{El-Badry}, K., {Ostriker}, E.~C., {Kim}, C.-G., {Quataert}, E., \& {Weisz}, D.~R. 2019, \mnras, 490, 1961, \dodoi{10.1093/mnras/stz2773}

\bibitem[{{Evoli} {et~al.}(2018){Evoli}, {Blasi}, {Morlino}, \& {Aloisio}}]{2018PhRvL.121b1102E}
{Evoli}, C., {Blasi}, P., {Morlino}, G., \& {Aloisio}, R. 2018, \prl, 121, 021102, \dodoi{10.1103/PhysRevLett.121.021102}

\bibitem[{{Farmer} \& {Goldreich}(2004)}]{2004ApJ...604..671F}
{Farmer}, A.~J., \& {Goldreich}, P. 2004, \apj, 604, 671, \dodoi{10.1086/382040}

\bibitem[{{Felice} \& {Kulsrud}(2001)}]{2001ApJ...553..198F}
{Felice}, G.~M., \& {Kulsrud}, R.~M. 2001, \apj, 553, 198, \dodoi{10.1086/320651}

\bibitem[{{Ferrando} {et~al.}(1988){Ferrando}, {Webber}, {Goret}, {Kish}, {Schrier}, {Soutoul}, \& {Testard}}]{1988PhRvC..37.1490F}
{Ferrando}, P., {Webber}, W.~R., {Goret}, P., {et~al.} 1988, \prc, 37, 1490, \dodoi{10.1103/PhysRevC.37.1490}

\bibitem[{{Ferri{\`e}re}(2001)}]{2001RvMP...73.1031F}
{Ferri{\`e}re}, K.~M. 2001, Reviews of Modern Physics, 73, 1031, \dodoi{10.1103/RevModPhys.73.1031}

\bibitem[{{Fielding} \& {Bryan}(2022)}]{2022ApJ...924...82F}
{Fielding}, D.~B., \& {Bryan}, G.~L. 2022, \apj, 924, 82, \dodoi{10.3847/1538-4357/ac2f41}

\bibitem[{{Fielding} {et~al.}(2020){Fielding}, {Ostriker}, {Bryan}, \& {Jermyn}}]{2020ApJ...894L..24F}
{Fielding}, D.~B., {Ostriker}, E.~C., {Bryan}, G.~L., \& {Jermyn}, A.~S. 2020, \apjl, 894, L24, \dodoi{10.3847/2041-8213/ab8d2c}

\bibitem[{{Fisk}(1976)}]{1976JGR....81.4633F}
{Fisk}, L.~A. 1976, \jgr, 81, 4633, \dodoi{10.1029/JA081i025p04633}

\bibitem[{{Foote} \& {Kulsrud}(1979)}]{1979ApJ...233..302F}
{Foote}, E.~A., \& {Kulsrud}, R.~M. 1979, \apj, 233, 302, \dodoi{10.1086/157391}

\bibitem[{{Ginzburg} {et~al.}(1973){Ginzburg}, {Ptuskin}, \& {Tsytovich}}]{1973Ap&SS..21...13G}
{Ginzburg}, V.~L., {Ptuskin}, V.~S., \& {Tsytovich}, V.~N. 1973, \apss, 21, 13, \dodoi{10.1007/BF00642191}

\bibitem[{{Girichidis} {et~al.}(2016){Girichidis}, {Naab}, {Walch}, {Hanasz}, {Mac Low}, {Ostriker}, {Gatto}, {Peters}, {W{\"u}nsch}, {Glover}, {Klessen}, {Clark}, \& {Baczynski}}]{2016ApJ...816L..19G}
{Girichidis}, P., {Naab}, T., {Walch}, S., {et~al.} 2016, \apjl, 816, L19, \dodoi{10.3847/2041-8205/816/2/L19}

\bibitem[{{Goldreich} \& {Sridhar}(1995)}]{1995ApJ...438..763G}
{Goldreich}, P., \& {Sridhar}, S. 1995, \apj, 438, 763, \dodoi{10.1086/175121}

\bibitem[{{Grappin} {et~al.}(1993){Grappin}, {Velli}, \& {Mangeney}}]{1993PhRvL..70.2190G}
{Grappin}, R., {Velli}, M., \& {Mangeney}, A. 1993, \prl, 70, 2190, \dodoi{10.1103/PhysRevLett.70.2190}

\bibitem[{{Grenier} {et~al.}(2015){Grenier}, {Black}, \& {Strong}}]{2015ARA&A..53..199G}
{Grenier}, I.~A., {Black}, J.~H., \& {Strong}, A.~W. 2015, \araa, 53, 199, \dodoi{10.1146/annurev-astro-082214-122457}

\bibitem[{{Gronke} \& {Oh}(2018)}]{Gronke2018}
{Gronke}, M., \& {Oh}, S.~P. 2018, \mnras, 480, L111, \dodoi{10.1093/mnrasl/sly131}

\bibitem[{{Guo} \& {Oh}(2008)}]{2008MNRAS.384..251G}
{Guo}, F., \& {Oh}, S.~P. 2008, \mnras, 384, 251, \dodoi{10.1111/j.1365-2966.2007.12692.x}

\bibitem[{{Haggerty} {et~al.}(2019){Haggerty}, {Caprioli}, \& {Zweibel}}]{2019ICRC...36..279H}
{Haggerty}, C., {Caprioli}, D., \& {Zweibel}, E. 2019, in International Cosmic Ray Conference, Vol.~36, 36th International Cosmic Ray Conference (ICRC2019), 279.
\newblock \doarXiv{1909.06346}

\bibitem[{{Hanasz} {et~al.}(2013){Hanasz}, {Lesch}, {Naab}, {Gawryszczak}, {Kowalik}, \& {W{\'o}lta{\'n}ski}}]{2013ApJ...777L..38H}
{Hanasz}, M., {Lesch}, H., {Naab}, T., {et~al.} 2013, \apjl, 777, L38, \dodoi{10.1088/2041-8205/777/2/L38}

\bibitem[{{Holcomb} \& {Spitkovsky}(2019)}]{2019ApJ...882....3H}
{Holcomb}, C., \& {Spitkovsky}, A. 2019, \apj, 882, 3, \dodoi{10.3847/1538-4357/ab328a}

\bibitem[{{Holman} {et~al.}(1979){Holman}, {Ionson}, \& {Scott}}]{1979ApJ...228..576H}
{Holman}, G.~D., {Ionson}, J.~A., \& {Scott}, J.~S. 1979, \apj, 228, 576, \dodoi{10.1086/156881}

\bibitem[{{Hopkins} {et~al.}(2022){Hopkins}, {Squire}, {Butsky}, \& {Ji}}]{2022MNRAS.517.5413H}
{Hopkins}, P.~F., {Squire}, J., {Butsky}, I.~S., \& {Ji}, S. 2022, \mnras, 517, 5413, \dodoi{10.1093/mnras/stac2909}

\bibitem[{{Hopkins} {et~al.}(2021){Hopkins}, {Squire}, {Chan}, {Quataert}, {Ji}, {Kere{\v{s}}}, \& {Faucher-Gigu{\`e}re}}]{2021MNRAS.501.4184H}
{Hopkins}, P.~F., {Squire}, J., {Chan}, T.~K., {et~al.} 2021, \mnras, 501, 4184, \dodoi{10.1093/mnras/staa3691}

\bibitem[{{Hu} \& {Krommes}(1994)}]{1994PhPl....1..863H}
{Hu}, G., \& {Krommes}, J.~A. 1994, Physics of Plasmas, 1, 863, \dodoi{10.1063/1.870745}

\bibitem[{{Iroshnikov}(1964)}]{1964SvA.....7..566I}
{Iroshnikov}, P.~S. 1964, \sovast, 7, 566

\bibitem[{{Jacob} {et~al.}(2018){Jacob}, {Pakmor}, {Simpson}, {Springel}, \& {Pfrommer}}]{2018MNRAS.475..570J}
{Jacob}, S., {Pakmor}, R., {Simpson}, C.~M., {Springel}, V., \& {Pfrommer}, C. 2018, \mnras, 475, 570, \dodoi{10.1093/mnras/stx3221}

\bibitem[{{Ji} {et~al.}(2019){Ji}, {Oh}, \& {Masterson}}]{2019MNRAS.487..737J}
{Ji}, S., {Oh}, S.~P., \& {Masterson}, P. 2019, \mnras, 487, 737, \dodoi{10.1093/mnras/stz1248}

\bibitem[{{Jiang} \& {Oh}(2018)}]{2018ApJ...854....5J}
{Jiang}, Y.-F., \& {Oh}, S.~P. 2018, \apj, 854, 5, \dodoi{10.3847/1538-4357/aaa6ce}

\bibitem[{{Jokipii}(1966)}]{1966ApJ...146..480J}
{Jokipii}, J.~R. 1966, \apj, 146, 480, \dodoi{10.1086/148912}

\bibitem[{{Kim} {et~al.}(2023){Kim}, {Gong}, {Kim}, \& {Ostriker}}]{Kim2023}
{Kim}, J.-G., {Gong}, M., {Kim}, C.-G., \& {Ostriker}, E.~C. 2023, \apjs, 264, 10, \dodoi{10.3847/1538-4365/ac9b1d}

\bibitem[{{Klein} {et~al.}(1994){Klein}, {McKee}, \& {Colella}}]{1994ApJ...420..213K}
{Klein}, R.~I., {McKee}, C.~F., \& {Colella}, P. 1994, \apj, 420, 213, \dodoi{10.1086/173554}

\bibitem[{{Kolmogorov}(1941)}]{1941DoSSR..30..301K}
{Kolmogorov}, A. 1941, Akademiia Nauk SSSR Doklady, 30, 301

\bibitem[{{Kraichnan}(1965)}]{1965PhFl....8.1385K}
{Kraichnan}, R.~H. 1965, Physics of Fluids, 8, 1385, \dodoi{10.1063/1.1761412}

\bibitem[{{Kulsrud} \& {Pearce}(1969)}]{1969ApJ...156..445K}
{Kulsrud}, R., \& {Pearce}, W.~P. 1969, \apj, 156, 445, \dodoi{10.1086/149981}

\bibitem[{{Kunz} {et~al.}(2014){Kunz}, {Stone}, \& {Bai}}]{2014JCoPh.259..154K}
{Kunz}, M.~W., {Stone}, J.~M., \& {Bai}, X.-N. 2014, Journal of Computational Physics, 259, 154, \dodoi{10.1016/j.jcp.2013.11.035}

\bibitem[{{Lancaster} {et~al.}(2024){Lancaster}, {Ostriker}, {Kim}, {Kim}, \& {Bryan}}]{Lancaster2024}
{Lancaster}, L., {Ostriker}, E.~C., {Kim}, C.-G., {Kim}, J.-G., \& {Bryan}, G.~L. 2024, \apj, 970, 18, \dodoi{10.3847/1538-4357/ad47f6}

\bibitem[{{Lancaster} {et~al.}(2021{\natexlab{a}}){Lancaster}, {Ostriker}, {Kim}, \& {Kim}}]{Lancaster2021}
{Lancaster}, L., {Ostriker}, E.~C., {Kim}, J.-G., \& {Kim}, C.-G. 2021{\natexlab{a}}, \apj, 914, 89, \dodoi{10.3847/1538-4357/abf8ab}

\bibitem[{{Lancaster} {et~al.}(2021{\natexlab{b}}){Lancaster}, {Ostriker}, {Kim}, \& {Kim}}]{2021ApJ...914...90L}
---. 2021{\natexlab{b}}, \apj, 914, 90, \dodoi{10.3847/1538-4357/abf8ac}

\bibitem[{{Lazarian}(2016)}]{2016ApJ...833..131L}
{Lazarian}, A. 2016, \apj, 833, 131, \dodoi{10.3847/1538-4357/833/2/131}

\bibitem[{{Lazarian} \& {Beresnyak}(2006)}]{2006MNRAS.373.1195L}
{Lazarian}, A., \& {Beresnyak}, A. 2006, \mnras, 373, 1195, \dodoi{10.1111/j.1365-2966.2006.11093.x}

\bibitem[{{Lebiga} {et~al.}(2018){Lebiga}, {Santos-Lima}, \& {Yan}}]{2018MNRAS.476.2779L}
{Lebiga}, O., {Santos-Lima}, R., \& {Yan}, H. 2018, \mnras, 476, 2779, \dodoi{10.1093/mnras/sty309}

\bibitem[{{Lee} \& {V{\"o}lk}(1973)}]{1973Ap&SS..24...31L}
{Lee}, M.~A., \& {V{\"o}lk}, H.~J. 1973, \apss, 24, 31, \dodoi{10.1007/BF00648673}

\bibitem[{{Lemmerz} {et~al.}(2024){Lemmerz}, {Shalaby}, {Pfrommer}, \& {Thomas}}]{2024arXiv240604400L}
{Lemmerz}, R., {Shalaby}, M., {Pfrommer}, C., \& {Thomas}, T. 2024, arXiv e-prints, arXiv:2406.04400, \dodoi{10.48550/arXiv.2406.04400}

\bibitem[{{Lerche}(1967)}]{1967ApJ...147..689L}
{Lerche}, I. 1967, \apj, 147, 689, \dodoi{10.1086/149045}

\bibitem[{{Livadiotis} \& {McComas}(2013)}]{2013SSRv..175..183L}
{Livadiotis}, G., \& {McComas}, D.~J. 2013, \ssr, 175, 183, \dodoi{10.1007/s11214-013-9982-9}

\bibitem[{{Lucek} \& {Bell}(2000)}]{2000MNRAS.314...65L}
{Lucek}, S.~G., \& {Bell}, A.~R. 2000, \mnras, 314, 65, \dodoi{10.1046/j.1365-8711.2000.03363.x}

\bibitem[{{Mandelker} {et~al.}(2020){Mandelker}, {Nagai}, {Aung}, {Dekel}, {Birnboim}, \& {van den Bosch}}]{2020MNRAS.494.2641M}
{Mandelker}, N., {Nagai}, D., {Aung}, H., {et~al.} 2020, \mnras, 494, 2641, \dodoi{10.1093/mnras/staa812}

\bibitem[{{McKenzie} \& {Voelk}(1982)}]{1982A&A...116..191M}
{McKenzie}, J.~F., \& {Voelk}, H.~J. 1982, \aap, 116, 191

\bibitem[{{Miller} {et~al.}(1996){Miller}, {Larosa}, \& {Moore}}]{1996ApJ...461..445M}
{Miller}, J.~A., {Larosa}, T.~N., \& {Moore}, R.~L. 1996, \apj, 461, 445, \dodoi{10.1086/177072}

\bibitem[{{Naab} \& {Ostriker}(2017)}]{2017ARA&A..55...59N}
{Naab}, T., \& {Ostriker}, J.~P. 2017, \araa, 55, 59, \dodoi{10.1146/annurev-astro-081913-040019}

\bibitem[{{Parker} \& {Lee}(1993)}]{1993PhFlB...5...77P}
{Parker}, S.~E., \& {Lee}, W.~W. 1993, Physics of Fluids B, 5, 77, \dodoi{10.1063/1.860870}

\bibitem[{{Pfrommer} {et~al.}(2017){Pfrommer}, {Pakmor}, {Schaal}, {Simpson}, \& {Springel}}]{2017MNRAS.465.4500P}
{Pfrommer}, C., {Pakmor}, R., {Schaal}, K., {Simpson}, C.~M., \& {Springel}, V. 2017, \mnras, 465, 4500, \dodoi{10.1093/mnras/stw2941}

\bibitem[{{Plotnikov} {et~al.}(2021){Plotnikov}, {Ostriker}, \& {Bai}}]{2021ApJ...914....3P}
{Plotnikov}, I., {Ostriker}, E.~C., \& {Bai}, X.-N. 2021, \apj, 914, 3, \dodoi{10.3847/1538-4357/abf7b3}

\bibitem[{{Ptuskin}(1997)}]{1997AdSpR..19..697P}
{Ptuskin}, V.~S. 1997, Advances in Space Research, 19, 697, \dodoi{10.1016/S0273-1177(97)00390-6}

\bibitem[{{Quataert} {et~al.}(2022){Quataert}, {Jiang}, \& {Thompson}}]{2022MNRAS.510..920Q}
{Quataert}, E., {Jiang}, Y.-F., \& {Thompson}, T.~A. 2022, \mnras, 510, 920, \dodoi{10.1093/mnras/stab3274}

\bibitem[{{Radin} {et~al.}(1974){Radin}, {Smith}, \& {Little}}]{1974PhRvC...9.1718R}
{Radin}, J.~R., {Smith}, A.~R., \& {Little}, N. 1974, \prc, 9, 1718, \dodoi{10.1103/PhysRevC.9.1718}

\bibitem[{{Reville} {et~al.}(2007){Reville}, {Kirk}, {Duffy}, \& {O'Sullivan}}]{2007A&A...475..435R}
{Reville}, B., {Kirk}, J.~G., {Duffy}, P., \& {O'Sullivan}, S. 2007, \aap, 475, 435, \dodoi{10.1051/0004-6361:20078336}

\bibitem[{{Roe}(1981)}]{1981JCoPh..43..357R}
{Roe}, P.~L. 1981, Journal of Computational Physics, 43, 357, \dodoi{10.1016/0021-9991(81)90128-5}

\bibitem[{{Schlickeiser}(1989)}]{1989ApJ...336..243S}
{Schlickeiser}, R. 1989, \apj, 336, 243, \dodoi{10.1086/167009}

\bibitem[{{Schlickeiser}(2002)}]{2002cra..book.....S}
---. 2002, {Cosmic Ray Astrophysics} (Springer Science \& Business Media)

\bibitem[{{Schlickeiser} \& {Miller}(1998)}]{1998ApJ...492..352S}
{Schlickeiser}, R., \& {Miller}, J.~A. 1998, \apj, 492, 352, \dodoi{10.1086/305023}

\bibitem[{{Sironi} \& {Narayan}(2015)}]{2015ApJ...800...88S}
{Sironi}, L., \& {Narayan}, R. 2015, \apj, 800, 88, \dodoi{10.1088/0004-637X/800/2/88}

\bibitem[{{Skilling}(1971)}]{1971ApJ...170..265S}
{Skilling}, J. 1971, \apj, 170, 265, \dodoi{10.1086/151210}

\bibitem[{{Skilling}(1975{\natexlab{a}})}]{1975MNRAS.173..255S}
---. 1975{\natexlab{a}}, \mnras, 173, 255, \dodoi{10.1093/mnras/173.2.255}

\bibitem[{{Skilling}(1975{\natexlab{b}})}]{1975MNRAS.173..245S}
---. 1975{\natexlab{b}}, \mnras, 173, 245, \dodoi{10.1093/mnras/173.2.245}

\bibitem[{{Skilling}(1975{\natexlab{c}})}]{1975MNRAS.172..557S}
---. 1975{\natexlab{c}}, \mnras, 172, 557, \dodoi{10.1093/mnras/172.3.557}

\bibitem[{{Soler} {et~al.}(2016){Soler}, {Terradas}, {Oliver}, \& {Ballester}}]{2016A&A...592A..28S}
{Soler}, R., {Terradas}, J., {Oliver}, R., \& {Ballester}, J.~L. 2016, \aap, 592, A28, \dodoi{10.1051/0004-6361/201628722}

\bibitem[{{Squire} {et~al.}(2020){Squire}, {Chandran}, \& {Meyrand}}]{2020ApJ...891L...2S}
{Squire}, J., {Chandran}, B.~D.~G., \& {Meyrand}, R. 2020, \apjl, 891, L2, \dodoi{10.3847/2041-8213/ab74e1}

\bibitem[{{Squire} {et~al.}(2021){Squire}, {Hopkins}, {Quataert}, \& {Kempski}}]{2021MNRAS.502.2630S}
{Squire}, J., {Hopkins}, P.~F., {Quataert}, E., \& {Kempski}, P. 2021, \mnras, 502, 2630, \dodoi{10.1093/mnras/stab179}

\bibitem[{{Stone} \& {Gardiner}(2009)}]{2009NewA...14..139S}
{Stone}, J.~M., \& {Gardiner}, T. 2009, \na, 14, 139, \dodoi{10.1016/j.newast.2008.06.003}

\bibitem[{{Stone} {et~al.}(2008){Stone}, {Gardiner}, {Teuben}, {Hawley}, \& {Simon}}]{2008ApJS..178..137S}
{Stone}, J.~M., {Gardiner}, T.~A., {Teuben}, P., {Hawley}, J.~F., \& {Simon}, J.~B. 2008, \apjs, 178, 137, \dodoi{10.1086/588755}

\bibitem[{{Stone} {et~al.}(2020){Stone}, {Tomida}, {White}, \& {Felker}}]{2020ApJS..249....4S}
{Stone}, J.~M., {Tomida}, K., {White}, C.~J., \& {Felker}, K.~G. 2020, \apjs, 249, 4, \dodoi{10.3847/1538-4365/ab929b}

\bibitem[{{Summers} \& {Thorne}(1991)}]{1991PhFlB...3.1835S}
{Summers}, D., \& {Thorne}, R.~M. 1991, Physics of Fluids B, 3, 1835, \dodoi{10.1063/1.859653}

\bibitem[{{Sun} \& {Bai}(2023)}]{2023MNRAS.523.3328S}
{Sun}, X., \& {Bai}, X.-N. 2023, \mnras, 523, 3328, \dodoi{10.1093/mnras/stad1548}

\bibitem[{{Tan} {et~al.}(2021){Tan}, {Oh}, \& {Gronke}}]{Tan2021}
{Tan}, B., {Oh}, S.~P., \& {Gronke}, M. 2021, \mnras, 502, 3179, \dodoi{10.1093/mnras/stab053}

\bibitem[{{Taylor}(1997)}]{1997ieas.book.....T}
{Taylor}, J. 1997, {Introduction to Error Analysis, the Study of Uncertainties in Physical Measurements, 2nd Edition}

\bibitem[{{Thomas} \& {Pfrommer}(2019)}]{2019MNRAS.485.2977T}
{Thomas}, T., \& {Pfrommer}, C. 2019, \mnras, 485, 2977, \dodoi{10.1093/mnras/stz263}

\bibitem[{{Tsung} {et~al.}(2022){Tsung}, {Oh}, \& {Jiang}}]{2022MNRAS.513.4464T}
{Tsung}, T. H.~N., {Oh}, S.~P., \& {Jiang}, Y.-F. 2022, \mnras, 513, 4464, \dodoi{10.1093/mnras/stac1123}

\bibitem[{{Uhlig} {et~al.}(2012){Uhlig}, {Pfrommer}, {Sharma}, {Nath}, {En{\ss}lin}, \& {Springel}}]{2012MNRAS.423.2374U}
{Uhlig}, M., {Pfrommer}, C., {Sharma}, M., {et~al.} 2012, \mnras, 423, 2374, \dodoi{10.1111/j.1365-2966.2012.21045.x}

\bibitem[{{Voelk}(1975)}]{1975RvGSP..13..547V}
{Voelk}, H.~J. 1975, Reviews of Geophysics and Space Physics, 13, 547, \dodoi{10.1029/RG013i004p00547}

\bibitem[{{V{\"o}lk}(1973)}]{volk1973}
{V{\"o}lk}, H.~J. 1973, Astrophysics and Space Science, 25, 471, \dodoi{10.1007/BF00649186}

\bibitem[{{Wentzel}(1968)}]{1968ApJ...152..987W}
{Wentzel}, D.~G. 1968, \apj, 152, 987, \dodoi{10.1086/149611}

\bibitem[{{Wentzel}(1969)}]{1969ApJ...156..303W}
---. 1969, \apj, 156, 303, \dodoi{10.1086/149965}

\bibitem[{{Wentzel}(1974)}]{1974ARA&A..12...71W}
---. 1974, \araa, 12, 71, \dodoi{10.1146/annurev.aa.12.090174.000443}

\bibitem[{{Wiener} {et~al.}(2013){Wiener}, {Oh}, \& {Guo}}]{2013MNRAS.434.2209W}
{Wiener}, J., {Oh}, S.~P., \& {Guo}, F. 2013, \mnras, 434, 2209, \dodoi{10.1093/mnras/stt1163}

\bibitem[{{Wiener} {et~al.}(2017{\natexlab{a}}){Wiener}, {Oh}, \& {Zweibel}}]{2017MNRAS.467..646W}
{Wiener}, J., {Oh}, S.~P., \& {Zweibel}, E.~G. 2017{\natexlab{a}}, \mnras, 467, 646, \dodoi{10.1093/mnras/stx109}

\bibitem[{{Wiener} {et~al.}(2017{\natexlab{b}}){Wiener}, {Pfrommer}, \& {Oh}}]{2017MNRAS.467..906W}
{Wiener}, J., {Pfrommer}, C., \& {Oh}, S.~P. 2017{\natexlab{b}}, \mnras, 467, 906, \dodoi{10.1093/mnras/stx127}

\bibitem[{{Wiener} {et~al.}(2018){Wiener}, {Zweibel}, \& {Oh}}]{2018MNRAS.473.3095W}
{Wiener}, J., {Zweibel}, E.~G., \& {Oh}, S.~P. 2018, \mnras, 473, 3095, \dodoi{10.1093/mnras/stx2603}

\bibitem[{{Xu} \& {Lazarian}(2022)}]{2022ApJ...927...94X}
{Xu}, S., \& {Lazarian}, A. 2022, \apj, 927, 94, \dodoi{10.3847/1538-4357/ac4dfd}

\bibitem[{{Yan} \& {Lazarian}(2011)}]{2011ApJ...731...35Y}
{Yan}, H., \& {Lazarian}, A. 2011, \apj, 731, 35, \dodoi{10.1088/0004-637X/731/1/35}

\bibitem[{{Zachary} \& {Cohen}(1986)}]{1986JCoPh..66..469Z}
{Zachary}, A.~L., \& {Cohen}, B.~I. 1986, Journal of Computational Physics, 66, 469, \dodoi{10.1016/0021-9991(86)90076-8}

\bibitem[{{Zeng} {et~al.}(2024){Zeng}, {Bai}, \& {Sun}}]{CRSI2D}
{Zeng}, S., {Bai}, X.-N., \& {Sun}, X. 2024, in preparation

\bibitem[{{Zhao} \& {Bai}(2023)}]{2023MNRAS.526.4245Z}
{Zhao}, X., \& {Bai}, X.-N. 2023, \mnras, 526, 4245, \dodoi{10.1093/mnras/stad3011}

\bibitem[{{Zweibel}(2013)}]{2013PhPl...20e5501Z}
{Zweibel}, E.~G. 2013, Physics of Plasmas, 20, 055501, \dodoi{10.1063/1.4807033}

\bibitem[{{Zweibel}(2017)}]{2017PhPl...24e5402Z}
---. 2017, Physics of Plasmas, 24, 055402, \dodoi{10.1063/1.4984017}

\bibitem[{{Zweibel}(2020)}]{2020ApJ...890...67Z}
---. 2020, \apj, 890, 67, \dodoi{10.3847/1538-4357/ab67bf}

\end{thebibliography}
\bibliographystyle{aasjournal}

\appendix
\section{Benchmark tests on for the adaptive $\delta f$ method}
We conduct two benchmark tests to validate the accuracy and performance enhancement of the adaptive $\delta f$ method (Section~\ref{sec::adapt_delta_f}). These benchmark problems closely match the simulation setup in this study, wherein the expanding box continuously drives the CR anisotropy, and the fitting method for $f_0$ remains consistent with Equation~\ref{equ::adapt_delta_f_fit_xi}~\&~\ref{equ::adapt_delta_f_fit_p_0}. However, the parameters in the subsequent simulations are simplified for benchmark purposes and do not hold astrophysical meanings.

\subsection{Accuracy test: fitting method for $f_0$ in the expanding box}
\label{sec::app_accuracy}
We simulate the adiabatic evolution of the CR population in the expanding box to validate the fitting method outlined by Equation~\ref{equ::adapt_delta_f_fit_xi}~\&~\ref{equ::adapt_delta_f_fit_p_0}. In the absence of isotropization from waves and solely driven by box expansion/compression, each CR particle conserves $p_\parallel a^2(t) = p\mu a^2(t)$ and $p_\perp a(t) = p a(t) \sqrt{1-\mu^2}$ (Equation~\ref{equ::pic}, also see the Appendix of \citealp{2023MNRAS.523.3328S}). Consequently, the CR distribution $f$, starting from an initial isotropic $\kappa$ distribution (Equation \ref{equ::kappa_dist_iso}), should conform to the bi-$\kappa$ distribution \citep{1991PhFlB...3.1835S},
\begin{align}
	f =&\frac{\rho_{\text{CR}}}{m \left(\pi \kappa p_0^2\right)^{1.5}} \frac{\mathcal{G}\left(\kappa + 1\right)}{\mathcal{G}\left(\kappa - 0.5 \right)} \bigg[1 + \frac{\left(p_\parallel a^2(t)\right)^2}{\kappa p_0^2} + \frac{\left(p_\perp a(t) \right)^2}{\kappa p_0^2}\bigg]^{-\kappa - 1} \notag \\ 
    &= \frac{\rho_{\text{CR}}}{m \left(\pi \kappa p_0^2\right)^{1.5}} \frac{\mathcal{G}\left(\kappa + 1\right)}{\mathcal{G}\left(\kappa - 0.5 \right)} \bigg[(1 + \frac{p^2}{\kappa p_0^2}\left(a^4 \mu^2  - a^2 \mu^2 + a^2\right)\bigg]^{-\kappa - 1}.
\end{align}
with the anisotropy parameter $\xi=a(t)$. In addition, the CR number density in the expanding box should depend on $a(t)$ as $\int f \dd^3 \boldsymbol{p} \propto a^{-4}(t)$. This leads to the anisotropic $\kappa$ distribution as given in Equation~\ref{equ::kappa_dist_aniso}. In this benchmark problem, taking the advantage that $\xi=a(t)$ is accurately known in the absence of waves, we utilize Equations~\ref{equ::adapt_delta_f_fit_xi}~and~\ref{equ::adapt_delta_f_fit_p_0} to dynamically fit $\xi$ and the peak momentum $p_0$ and test the accuracy of our adaptive $\delta f$ method.

The simulation setup largely follows the fiducial runs in the expanding/compressing boxes, albeit slightly simplified for test purposes. We initialize CRs in a uniform isotropic $\kappa$ distribution (Equation \ref{equ::kappa_dist_iso}) with $p_0=200 U_A$ and $\kappa=1.25$. The simulation particles are segregated into eight momentum bins, with $2.4576\times 10^6$ particles allocated to each bin, matching the total number of simulation particles in the fiducial runs (Section \ref{sec::setup}). To prevent isotropization, we set the initial wave amplitudes to zero and disable the CR backreaction. The background field and the expanding box setup are identical to those in Section \ref{sec::setup}, while the expansion/compression rate is elevated to $\dot{a} = \pm 10^{-2} \Omega_0$. The simulation spans $10^2 \Omega_0^{-1}$, where the CR anisotropy becomes significant by the end of the simulation, with $f_0$ updated every $\Delta T_{\text{adapt}} = 1 \Omega_0^{-1}$.

\begin{figure}
	\includegraphics[width=\columnwidth]{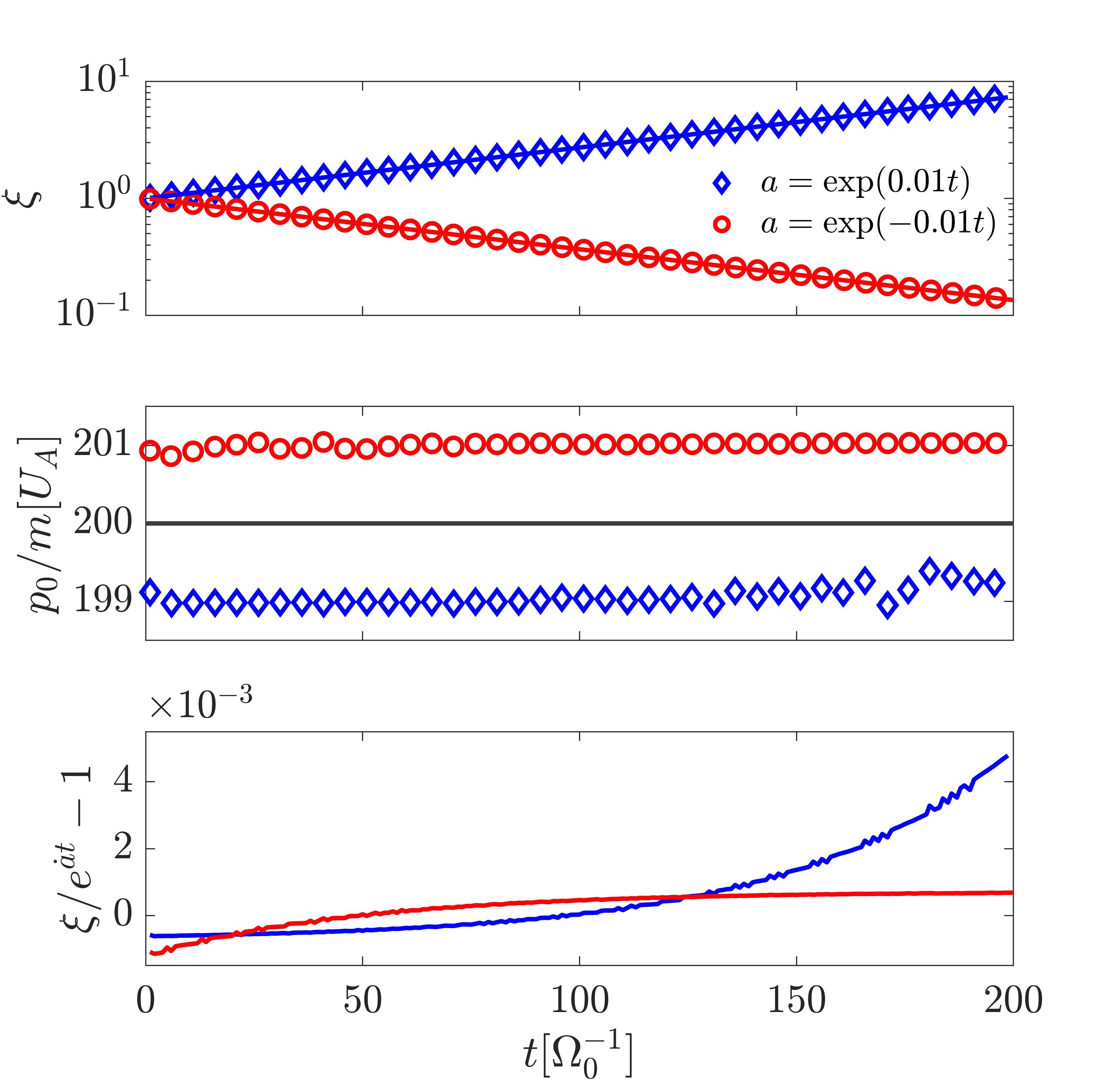}
	\caption{
		 Fitting results for the CR anisotropy parameter $\xi$ and the peak momentum $p_0$ in the test of the adaptive $\delta f$ method using expanding box (blue) and compressing box (red). The box expansion rate $\dot{a}$ is $\pm 10^{-2} \Omega_0$. The markers in the top two panels represent the fitting results obtained from Equation~\ref{equ::adapt_delta_f_fit_xi} and Equation~\ref{equ::adapt_delta_f_fit_p_0}, while the solid lines depict the expected values. The bottom panel illustrates the relative deviation of the fitted $\xi$ from the expected values.
	}
    \label{fig::app_accuracy}
\end{figure}

The fitting results for $\xi$ and $p_0$ are depicted in Figure~\ref{fig::app_accuracy}. The fitted $\xi$ closely tracks the expansion rate $a(t) = \exp(\dot{a}t)$, while the fitted $p_0$ remains constant around the initial value of $200 U_A$, in both the expanding box and the compressing box. We note that there is a systematic deviation in the fitted $p_0$ less than 1\% relative to the anticipated values. This small deviation is related to the initialized CR distribution, where initial simulation particles only span over a finite range of the $\kappa$ distribution ($p_0/500$ to $500 p_0$, see Section~\ref{sec::setup})\footnote{The fitting in the adaptive $\delta f$ method that we use here is derived from the generalized moments of the entire $\kappa$ distribution. The high-energy tail of the $\kappa$ distribution has a greater `weight' for high-order moments. Consequently, when the simulated distribution is truncated at $500p_0$, the measured $p_0$ from Equation~\ref{equ::adapt_delta_f_fit_p_0_full} will be smaller than the exact $p_0$ in the distribution, approximately by 10\% for $\kappa = 1.25$. To further correct the fitting error, we adopt Equation~\ref{equ::adapt_delta_f_fit_p_0}, which adjusts the fitting $p_0$ by the last-step $p_0$. Equation~\ref{equ::adapt_delta_f_fit_p_0} mitigates the fitting deviation on the order of $\delta f_\text{old} / f$. The expanding and compressing boxes lead the pitch angle distribution to evolve in opposite directions. Therefore, $\delta f_\text{old}$ has opposite signs in the expanding and compressing boxes, resulting in a positive error for $p_0$ in the compressing box and a negative error in the expanding box.}. The fitting error can be further minimized by covering over a larger range of the $\kappa$ distribution or increasing the fitting frequency  (decreasing $\Delta T_{\text{adapt}} \dot{a}$).

\subsection{Performance test: signal-to-noise improvement in CPPAI}
\label{sec::app_performance}

We compare the wave spectra triggered by CRPAI in the expanding box, both with the adaptive $\delta f$ method or the traditional $\delta f$ method, to assess the effectiveness. The adaptive $\delta f$ method can alleviate the statistical noise in the CR backreaction, particularly in scenarios where the CR distribution $f$ is less predictable.

The simulation setup closely resembles that in Section~\ref{sec::setup}, albeit with simplifications tailored for testing purposes. In comparison to the fiducial runs, we eliminate the ion-neutral damping and augment the box expansion rate to $\dot{a} = + 5 \times 10^{-5} \Omega_0$, to expedite both the wave isotropization and the anisotropy driving. Additionally, for a significant wave intensity within a limited runtime, we set the initial wave amplitude as $A=10^{-2}$ and enhance the CR mass ratio to $\rho_{\text{CR}}/\rho_0=10^{-4}$. We set the 1D simulation box size to $L_x=9.6 \times 10^4 U_A/\Omega_0$ and the resolution to $10 U_A/\Omega_0$. All other parameters and setups, such as the initial CR distribution and the number of momentum bins for initial simulation particles, remain identical to those of the fiducial run in the expanding box.

\begin{figure}
	\includegraphics[width=\columnwidth]{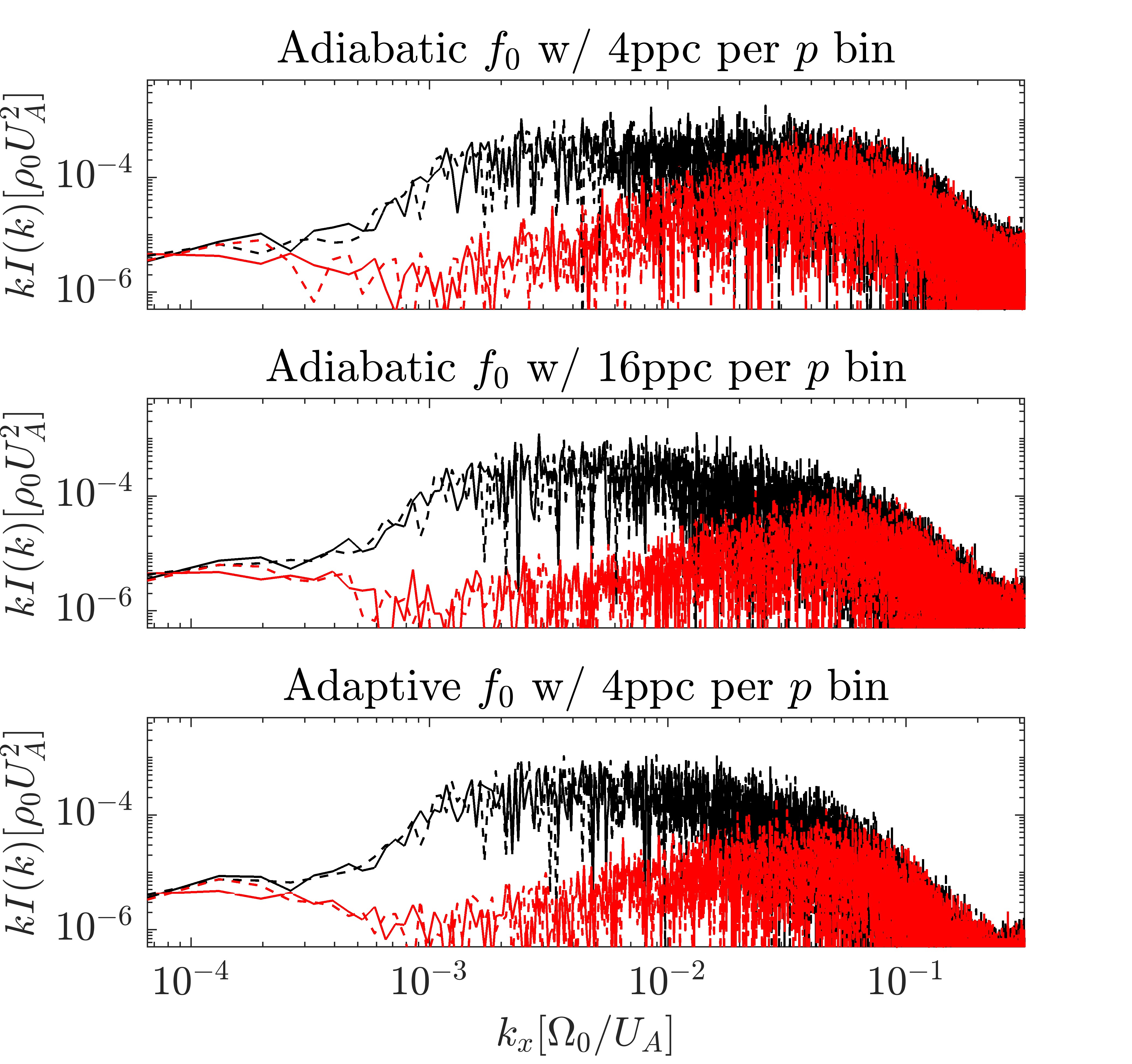}
	\caption{
		Wave intensity spectra, $kI(k)$, in the benchmark problem with varying $f_0$ and numbers of simulation particles, at time $t=3 \times 10^4 \Omega_0^{-1}$. The color and line styles correspond to those in Figure~\ref{fig::fid_wave_spec}. Lines of the same color (representing polarization) substantially overlap.
	}
    \label{fig::app_performance}
\end{figure}

We carry out three parallel runs by varying the choice $f_0$ in the $\delta f$ method and particles per cell per momentum bin: the adaptive $f_0$ with four particles per cell (ppc) per momentum bin, the adiabatic $f_0$ with four ppc per momentum bin, and the adiabatic $f_0$ with 16 ppc per momentum bin. The adaptive $f_0$ fits $\xi$ and $p_0$ every $\Delta T_{\text{adapt}} = 5\Omega_0^{-1}$ (see Section~\ref{sec::adapt_delta_f}). The adiabatic $f_0$ maintains $\xi = a(t)$ and a constant $p_0$, representing CR adiabatic evolution without wave isotropization. We illustrate the wave spectra measured at $t=3 \times 10^4 \Omega_0^{-1}$ in Figure~\ref{fig::app_performance}. Comparing the spectra from the two adiabatic $f_0$ runs with different particles per cell, the statistical noise from the CR backreaction primarily affects the high-$k$ regime, and the run with 16ppc has a significantly smaller noise level. The adaptive $\delta f$ method even more effectively reduces the noise at high-$k$. With only 4ppc, the noise level is even smaller than the run with adiabatic $f_0$ but using four times more simulation particles.

The effectiveness of the adaptive $\delta f$ method depends on the specific problem at hand. In this benchmark, it significantly reduces computational costs by over four times compared to the traditional $\delta f$ method, where $f_0$ remains the one evolving adiabatically from the initial CR distribution. However, when $f$ is predictable, such as when isotropization from waves dominates, or when quasi-linear diffusion is negligible and CR evolution is adiabatic, the adaptive $\delta f$ method offers no significant improvement in signal-to-noise ratio. In our study, the saturation of CRPAI, where $f$ deviates from both isotropic distribution and adiabatic distribution, the adaptive $\delta f$ method proves essential. Note that the anisotropy level in the actual simulations varies with $p$, and further improvement may be considered by allowing $\xi$ to be a parameterized function of $p$, but this can be left for future work.



\end{document}